%% file: Performance_White_Paper.tex
\pdfoutput=1
\documentclass[runningheads]{llncs}
\usepackage{graphicx}
\usepackage{subcaption}
\usepackage{caption}
\usepackage{amsmath}
\usepackage{hyperref} % puts box around url (highlights)
\usepackage{xcolor}
\usepackage{multirow}% http://ctan.org/pkg/multirow
\usepackage{listings}
\usepackage{pgfplots}
\usepackage{amssymb}
\usepackage{pifont}% http://ctan.org/pkg/pifont
\newcommand{\cmark}{\ding{51}}%
\newcommand{\xmark}{\ding{55}}%
\usepackage{array}
\newcolumntype{P}[1]{>{\centering\arraybackslash}p{#1}}
\colorlet{punct}{red!60!black}
\definecolor{background}{HTML}{EEEEEE}
\definecolor{delim}{RGB}{20,105,176}
\colorlet{numb}{magenta!60!black}

\lstdefinelanguage{json}{
    basicstyle=\scriptsize\ttfamily,
    numbers=left,
    numberstyle=\scriptsize,
    stepnumber=1,
    numbersep=8pt,
    showstringspaces=false,
    breaklines=true,
    frame=lines,
    backgroundcolor=\color{background},
    literate=
     *{0}{{{\color{numb}0}}}{1}
      {1}{{{\color{numb}1}}}{1}
      {2}{{{\color{numb}2}}}{1}
      {3}{{{\color{numb}3}}}{1}
      {4}{{{\color{numb}4}}}{1}
      {5}{{{\color{numb}5}}}{1}
      {6}{{{\color{numb}6}}}{1}
      {7}{{{\color{numb}7}}}{1}
      {8}{{{\color{numb}8}}}{1}
      {9}{{{\color{numb}9}}}{1}
      {:}{{{\color{punct}{:}}}}{1}
      {,}{{{\color{punct}{,}}}}{1}
      {\{}{{{\color{delim}{\{}}}}{1}
      {\}}{{{\color{delim}{\}}}}}{1}
      {[}{{{\color{delim}{[}}}}{1}
      {]}{{{\color{delim}{]}}}}{1},
}

\begin{document}
%
%\title{These Go to Eleven: Scaling Enterprise Blockchain Applications by $100\times$}
%\title{These Go to Eleven: Performance Tuning and Scaling Enterprise Blockchain Applications}
\title{Performance Tuning and Scaling Enterprise Blockchain Applications}

%\titlerunning{Scaling Enterprise Blockchain Applications by $100\times$}
\titlerunning{Performance Tuning and Scaling Enterprise Blockchain Applications}
\author{
Grant Chung\inst{1} \and
Luc Desrosiers\inst{2} \and
Manav Gupta\inst{3} \and
Andrew Sutton\inst{1} \and
Kaushik Venkatadri\inst{1} \and
Ontak Wong\inst{1} \and
Goran Zugic\inst{1}}
\authorrunning{Chung G., Desrosiers L., Gupta M., Sutton A., et al.}
\institute{Royal Bank of Canada, Toronto ON, Canada
\email{\{grant.chung,andrew.sutton,kaushik.venkatadri,ontak.wong,goran.z.zugic\}@rbc.com} \and
IBM Blockchain Labs, Hursley, UK\\
\email{ldesrosi@uk.ibm.com} \and
IBM Toronto Software Lab, Markham ON, Canada\\
\email{manavgup@ca.ibm.com}}
\maketitle              % typeset the header of the contribution
\begin{abstract}
Blockchain scalability can be complicated and costly. As enterprises begin to adopt blockchain technology to solve business problems, there are valid concerns if blockchain applications can support the transactional demands of production systems. In fact, the multiple distributed components and protocols that underlie blockchain applications makes performance optimization a non-trivial task. Blockchain performance optimization and scalability require a methodology to reduce complexity and cost. Furthermore, existing performance results often lack the requirements, load, and infrastructure of a production application. In this paper, we first develop a methodical approach to performance tuning enterprise blockchain applications to increase performance and transaction capacity. The methodology is applied to an enterprise blockchain-based application (leveraging Hyperledger Fabric) for performance tuning and optimization with the goal of bridging the gap between laboratory and production deployed system performance. We then present extensive results and analysis of our performance testing for on-premise and cloud deployments, in which we were able to scale the application from 30 to 3000 TPS without forking the Hyperledger Fabric source code and maintaining a reasonable infrastructure footprint. We also provide blockchain application and platform recommendations for performance improvement.
%The abstract should briefly summarize the contents of the paper in

\end{abstract}

\input{Sections/Introduction}

\input{Sections/Related_Work}

\input{Sections/System_Overview}

\input{Sections/Performance_Optimization_Methodology}

\input{Sections/Experiment_Setup}

\input{Sections/Tunable_Components}

\input{Sections/Experiment_Summary}

\input{Sections/Component_Scaling}

\input{Sections/Specialty_Tests}

\input{Sections/Cloud_Deployment}

\input{Sections/Improvements}

\input{Sections/Conclusion}

\bibliographystyle{splncs04}
\bibliography{biblio}

\end{document}

%% file: Sections/Introduction.tex
\section{Introduction}

Distributed systems are prevalent in modern computing platforms \cite{wattenhofer2019}. Many organizations are geographically distributed and require a system to support this dispersion of resources. Multi-core processors and computing clusters offer higher degrees of parallelism to further speed up computations compared to single core machines. Security threats such as ransomware (e.g., WannaCry \cite{wannacry}) reinforce the importance of data replication across multiple machines to prevent data loss (i.e., reliability) and to allow data access at any time (i.e., availability).

For the past decade, a synthesis of existing techniques from distributed systems and cryptography gave rise to blockchain technology: a distributed and decentralized solution for securely storing and executing transactions. Popularized by Bitcoin \cite{bitcoin}, and generalized by platforms such as Ethereum \cite{etheruem} and Hyperledger Fabric \cite{androulaki2018}, blockchain technology has grabbed the attention of enterprises for its capability to deliver proof of transactions in a decentralized network. This capability can eliminate costly point-to-point integration, and provide near real-time views of transacted assets and a single source of truth. Additionally, built-in data integrity (e.g., immutable transaction records) and non-repudiation  guarantees (e.g., digital signatures) 
%(e.g., immutable transaction records and permissioned networks) 
combined with data replication and high availability make blockchain technology an enticing solution for business problems. 

As enterprises look to adopt blockchain technology to solve their business problems, there is hesitation to fully deploy blockchain-based solutions to production environments due to the misconstrued notion that blockchain technology does not scale. For example, Bitcoin is limited to an average of 7 transactions per second (TPS), which severely impacts scalability  \cite{croman2016}. Improvements to the underlying platforms (e.g., consensus protocols such as HotStuff \cite{hotstuff}) and the emergence of permissioned networks (e.g., Hyperledger Fabric) have shown that blockchain-based systems can reach high throughput \cite{ibmindia,fastfabric}. However, many of these results are reported from lab settings and lack the application requirements, load, and infrastructure of a production application. For example, there are vast differences in reported throughput for the Hyperledger Fabric platform (e.g., \cite{androulaki2018,ibmindia,nystrom2019,nguyen2019}), which shows the difficulty of configuring and tuning distributed systems. This motivates the need for a systematic approach to performance tuning distributed systems, in particular blockchain systems, so that they can maximize performance potential. Additionally, the complexity of a distributed blockchain system often wrongfully detracts from the fundamental benefits the system provides over traditional approaches.

\begin{figure}[t]
\captionsetup{font=footnotesize}
    \centering
    \begin{subfigure}[b]{0.2\textwidth}
        \includegraphics[width=\textwidth]{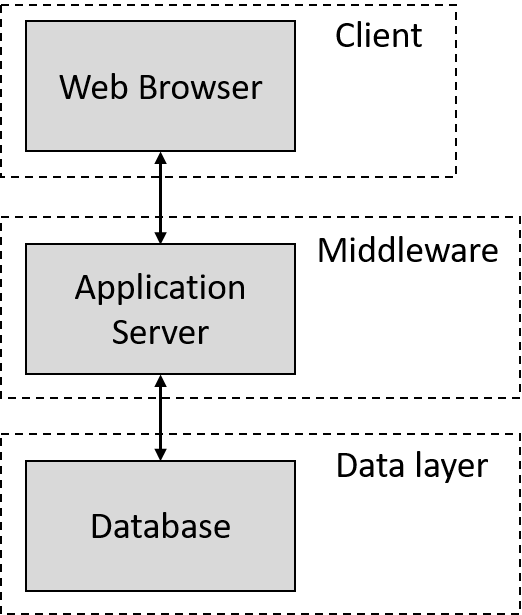}
        \caption{Classic Web Application}
        \label{fig:classic}
    \end{subfigure}
    ~ %add desired spacing between images, e. g. ~, \quad, \qquad, \hfill etc. 
    \begin{subfigure}[b]{0.7\textwidth}
        \includegraphics[width=\textwidth]{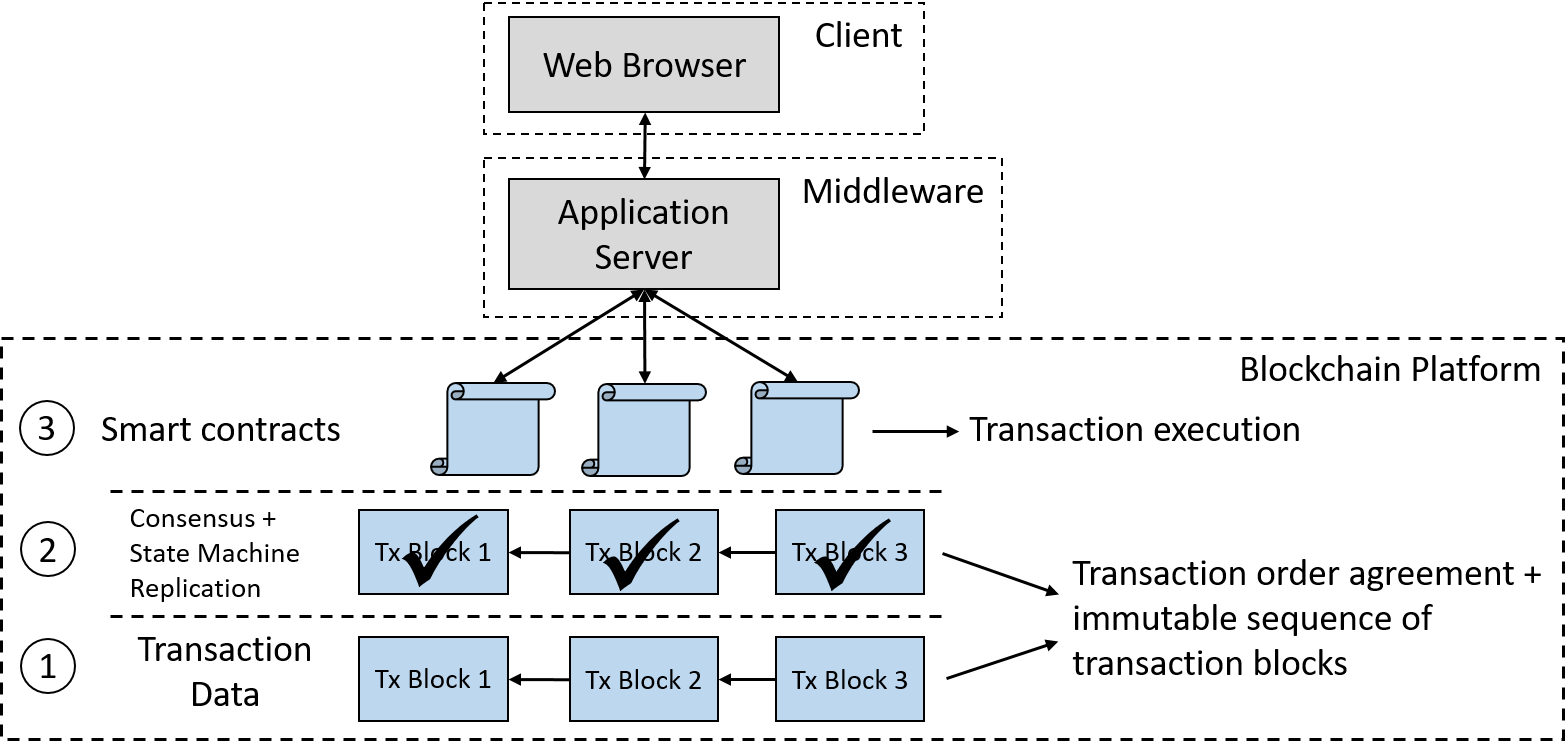}
        \caption{Blockchain-based Web Application \cite{abraham2017}}
        \label{fig:blockchain}
    \end{subfigure}
    \caption{Web Application Architectures}\label{fig:webapps}
\end{figure}

Figure \ref{fig:classic} illustrates an example of a typical enterprise 3-tiered web application architecture. A simplified transaction flow is: (i) the user submits a request through the client; (ii) transaction processing is done in the application server; (iii) data is read or updated in the data layer; (iv) and the response is sent back to the client. 
%Applications that follow this architecture pattern are typically centralized, both in terms of architecture and component ownership, and tracing a transaction through the system is relatively straight forward.
Alternatively, a distributed blockchain-based web application is illustrated in Figure \ref{fig:blockchain}. Compared to the classical web application, there is little difference from the user's perspective; requests are sent from the client and data is read or updated. However, from the backend's perspective, the application connects to a blockchain network, which can be decomposed into three layers \cite{abraham2017}. The data layer (1) stores an immutable sequence of transactions, which are batched into blocks, among participants who do not  necessarily trust each other. An underlying principle of blockchain is to achieve state replication across multiple nodes by having all participants execute the same set and sequence of transactions. This principle requires that participants come to agreement on the sequence of transactions, which is solved by the layer handling consensus (2). A consensus protocol is used to algorithmically agree on the order of transactions and form a chain of transaction blocks \cite{abraham2017}. The execution layer (3) exposes high-level business abstractions, known as smart contracts, over the shared data structure established in the first two layers \cite{abraham2017}. These contracts allow programmatically executing, verifying, and enforcing business transactions, which are agreed upon and replicated in the lower layers.

Blockchain-based web applications involve multiple distributed components and protocols working together that make performance optimization a non-trivial task. The complexity of managing the I/O, communication, computation, and the coordination between components contributes to this complexity. As shown in Figure \ref{fig:webapps}, compared to the classical web application, tracing a transaction through a blockchain application is more difficult since it passes through phases such as smart contract execution and consensus. To add to the complexity, these phases run on different components that can potentially be owned and operated by different organizations. Therefore, it is crucial to mitigate complexity's effect on performance by tuning and optimizing the system correctly.

The gap that this work intends to fill is to demonstrate how an enterprise blockchain application can be tuned and optimized in order to scale to meet growing transactional demands. First, we present a performance tuning methodology to systematically analyze complex distributed systems. Second, we apply this methodology to optimize a production-grade blockchain-based application that leverages the Hyperledger Fabric platform. We perform a number of experiments that leverage realistic transaction loads for enterprise use cases against an on-premise Z system and IBM Cloud deployed production blockchain application. The experiments include horizontally scaling application servers and peers, specialty tests (e.g., commit strategies), queries, infrastructure scaling (e.g., up to 24 CPU cores), and analysis of configurations, such as block size, the impact of scaling peers, and transaction phases. Based on the experimental results, the utilization of our performance optimization methodology scaled our on-premise and cloud deployed blockchain application from 30 to 1900 TPS and 1000 to 3000 TPS, respectively. Third, we provide recommendations to further improve performance, such as an asynchronous request handling design, threshold signatures, and application component distribution on the underlying infrastructure.

The key contributions of the paper are: (i) a methodology to tune and optimize a blockchain-based application; (ii) tuning parameters for application and blockchain platform optimizations; (iii) results and analysis from performance testing an enterprise blockchain application; and (iv) recommendations for further performance improvements. The rest of the paper is structured as follows. Section \ref{sec:relatedwork} provides an examination of related literature and the motivation for our work. In Section \ref{sec:overview}, we describe the design and components of our blockchain-based application. Section \ref{sec:methodology} outlines our performance optimization methodology. We describe the on-premise Z system infrastructure setup for our experiments and a summary of experiment results in Section \ref{sec:solution assess}. Section \ref{sec:params} describes the tunable application and system components that will be leveraged during the experiments. 
We then apply our optimization methodology to performance testing an enterprise blockchain application and report the extensive analysis of our baseline test, component scaling, and specialty test results in Sections \ref{sec:exp_baseline_summ}, \ref{sec:comp_scaling}, and \ref{sec:specialty}, respectively. Section \ref{sec:cloud} provides results for our IBM Cloud based performance testing. Performance improvement recommendations based on our experiment results are presented in Section \ref{sec:recommendations}.
%{\color{red} We investigate the tunable parameters that affect the performance of the system in Section \ref{sec:params}. 
%Section \ref{sec:results} shows the results and analysis of our performance tuning.} 
We provide future directions and conclude in Section \ref{sec:conlusion}.

%% file: Sections/Related_Work.tex
\section{Related Work} \label{sec:relatedwork}

In this section we investigate the following areas of related work: methodologies and frameworks, performance results and analysis, and improvements and optimizations.

\subsection{Methodologies \& Frameworks}

The multiple layers, components, and protocols of modern distributed systems contribute to the complexity of effective performance engineering; systematic methodologies help mitigate complexity's effect on performance. The common steps in systematic performance evaluations are described in \cite{jain1991} and generalized performance engineering approaches are presented in \cite{westermann2010,westermann2010b,woodside}.  Similar to our approach for enterprise applications, Westermann \textit{et al}.  propose model and measurement-based performance evaluation techniques for large enterprise applications \cite{westermann2010,westermann2010b}. 
%However, Westermann's approach requires each stakeholder developing plug-ins to integrate in the plaform 
%Similar to our approach for enterprise applications, Westermann et al. propose a model and measurement-based performance evaluation techniques for large enterprise applications \cite{westermann2010}. 
Woodside \textit{et al.} describe the software performance engineering domain and process concepts and survey the current approaches \cite{woodside}. The procedures introduced in these papers motivates our approach of using best practices in performance engineering, such as evaluation techniques and selecting performance metrics. We leverage some of these concepts to form our distributed system-centric performance optimization methodology.

There are several performance methodologies related to private blockchain platforms. Nasir \textit{et al.} provide a methodology for evaluating and assessing different blockchain platforms in terms of performance, security, and scalability \cite{nasir2018}. The methodology includes an  architecture for blockchain performance evaluation that consists of a performance analysis layer (modified Hyperledger Caliper \cite{caliper}), adapter layer (integration of blockchain implementations into Caliper), interface layer (deploy, invoke, and query smart contracts), and the blockchain platform (Hyperledger Fabric). Kocsis \textit{et al.} design a performance evaluation methodology for performance modeling blockchain technologies \cite{kocsis2017}. The performance model takes a measurement-based approach where bottleneck estimates are evaluated through targeted component sensitivity analysis. Pongnumkul \textit{et al.} describe a methodology for evaluating a blockchain platform based on pre-configured application simulations and evaluation workload dispatchers \cite{pongnumkul2017}. These methodologies do not provide the necessary depth for tuning complex distributed systems, such as enterprise blockchain applications, which makes determining cause-effect and component relationships difficult. There is also a lack of completeness in the requirements for analyzing a distributed system, such as the ability to trace transactions.

Frameworks extend methodologies by providing a structure for comparing and analyzing blockchain platforms. Dinh \textit{et al.} present the first blockchain evaluation framework, called Blockbench, for analyzing private blockchains \cite{blockbench}. The framework abstracts blockchain systems to four layers (consensus, data, execution, and application), in which analysis can be performed based on the distinct abstraction layers. Sukhwani describes models that provide a quantitative framework to help compare different deployment configurations of Hyperledger Fabric and make design trade-off decisions \cite{sukhwani2018}. By leveraging Stochastic Reward Nets, the framework captures system operations and complex interactions between components, and can compute performance for a system with proposed architectural improvements before they are implemented. Incorporating similar frameworks to our methodology would be beneficial for further insights into system component interactions. 
%Also it would help reduce the number of empirical tests to run and how to decide which test to run.

\subsection{Performance Results \& Analysis}

Hyperledger Fabric is a leading enterprise blockchain platform and is leveraged in this paper. Androulaki \textit{et al.} introduced Hyperledger Fabric  and reported preliminary performance results of 3500 TPS with sub-second latency using a data model based on unspent transaction output (UTXO) cryptocurrencies \cite{androulaki2018}. %(Fabric was not yet performance-tuned and optimized).
%Androulaki \textit{et al.} report that Fabric can achieve throughput of more than 3500 transactions per second with sub-second latency \cite{androulaki2018}. 
Thakkar \textit{et al.} \cite{ibmindia} report a peak throughput of about 850 TPS. Nystrøm incorporates network performance, including gossip network traffic, transaction message sizes, and network-based transaction flow data, with transaction throughput for performance measurement \cite{nystrom2019}. Reported throughput reached up to 1500 TPS. Nguyen \textit{et al.} \cite{nguyen2019} aim to conduct more realistic setups than existing performance experiments. With a 48-node commodity cluster with 32 GB RAM and 3.5GHz Xeon class CPUs, Nguyen \textit{et al.} report 400 TPS \cite{nguyen2019}. Baliga \textit{et al.} present an experimental approach to characterizing throughput and latency of Hyperledger Fabric \cite{baliga2018}.

Many of the work described above, e.g., \cite{nystrom2019,nguyen2019,sukhwani2018,nasir2018}, leverage benchmark frameworks such as Hyperledger Caliper \cite{caliper} to drive transactions to the network and report throughput results. However, using these frameworks bypasses system layers (such as the application tier) by interacting directly with blockchain peers and miss additional processing required in a production blockchain application. 
%For example, Hyperledger Caliper does not wait for transaction commit confirmations from peers (i.e., sends requests asynchronously). 
%This configuration is not representative of enterprise applications since in most use cases the client must have a guarantee of transaction commitment. Omitting the commit confirmation processing on the application tier provides significant performance improvements (Section \ref{sec:specialty}). 
There is also the omission of errors, where unsuccessful transaction executions (e.g., due to timeouts) contribute to inflated throughput results. 
For completeness, we leverage a full application deployment (client is simulated with JMeter) to include all additional application tier processing not included in many performance results.

The authors in \cite{androulaki2018} describe Fabric as a complex distributed system where performance depends on many parameters such as the choice of distributed application and transaction size, ordering service implementation, network parameters, node topology, number of nodes, endorsement policy, etc. This is further motivated by the vast differences in reported throughput from the work described above, where performance is highly dependent on deployment topology, available resources, tuning, and configuration. The difficulty in configuring and tuning a distributed system such as Hyperledger Fabric is a prime motivator for our work presented in this paper 
%methodology and realistic enterprise application performance testing. %You can see based on the vast throughput difference of reported results that there are many factors involved. 
and the importance of systematic analysis. Many performance results do not include the necessary load (e.g., number of concurrent users), complexity (e.g., application and smart contract implementation), or architecture and infrastructure requirements for enterprise use cases (e.g., commodity hardware \cite{androulaki2018}). For example, results do not take into account the infrastructure footprint (e.g., virtual machines running 32 GB RAM with high specification Intel Xeon CPUs \cite{nguyen2019}) or the transaction size is not representative of a number of real-world use cases (less than 4KB \cite{androulaki2018}). Our work aims to report more realistic results for enterprise applications in a production setting.

\subsection{Improvements \& Optimizations}

Generally, performance improvement and optimization results for Hyperledger Fabric fall into two categories: (i) architecture redesign and (ii) recommendations for reducing bottlenecks. There are several proposals and implementations for improving the scalability and performance of the Hyperledger Fabric platform by re-architecting the system. Gorenflo \textit{et al.} implement Hyperledger Fabric architectural optimizations to improve the end-to-end transaction throughput to 20,000 TPS \cite{fastfabric}. Javaid \textit{et al.} re-architect the validation phase for $2\times$ performance improvement \cite{haris2019}. Gorenflo \textit{et al.} introduce a hybrid execution model and concurrent transaction commitment to Fabric \cite{gorenflo2019}. % these papers implement changes

Performance bottlenecks of Hyperledger Fabric have been observed. Thakkar \textit{et al.} provided guidelines for configuring parameters for optimizing performance, identified platform bottlenecks, and proposed optimizations such as MSP cache, parallel transaction validation, and bulk read/write \cite{ibmindia}. In fact, these optimizations were incorporated in Hyperledger Fabric v1.1. 
%(we leverage v1.4.1). 
Sukhwani observed the implications of block size on transaction throughput and latency \cite{sukhwani2018}. % these papers recommend changes

Fundamentally, there are strong cases (e.g., \cite{fastfabric}) for redesigning aspects of the Fabric architecture to reduce the impact of bottlenecks in the system. %We fall into the latter category, but 
The optimization recommendations presented in this paper (Section \ref{sec:recommendations}) support and complement this body of performance work. Although redesigning some architectural aspects of Fabric may be required to achieve Visa-like throughput \cite{fastfabric}, our main objective is to demonstrate the effectiveness of applying a methodology to scaling a blockchain application based on current non-forked versions of blockchain platforms, rather than alter the underlying implementation. Additionally, we propose performance optimizations such as a buffered channel for validation, asynchronous request handling, and best practices for distributing components on the underlying infrastructure.

%% file: Sections/System_Overview.tex
\section{System Overview} \label{sec:overview}
In this section, we provide a high-level description of our blockchain-based web application. The business problem our system solves is described in Section \ref{sec:solution desc}.
%Figure \ref{fig:blockchain} illustrates the components in a blockchain-based application. 
Section \ref{sec:appserver} presents the middleware, which is deployed as a Node.js server, and the blockchain platform is described in Section \ref{sec:fabric}.  We omit the client-side UI in our performance tests since we simulate the client with JMeter, which is responsible for submitting transactions to the application through a REST gateway.
%we bypass the UI in our tests and interact with the application through a REST gateway. The client is simulated with JMeter, which is responsible for submitting transactions to the application.

\subsection{Solution Description} \label{sec:solution desc}

Back-office reconciliations are often sources of disagreement between business units of an organization. The lack of recorded evidence of the internal agreement between business units requires reconciliation, which is typically a manual process and not formalized. The solution in which we based our performance tuning and testing on eliminates the root causes for reconciliation by designing a new business process that can take advantage of blockchain technology. The core system is modeled around the notion of an agreement and every subsequent action on an agreement is carried out by consensus. Therefore, parties and counterparties to a transaction have to always agree or disagree together (i.e., it is not possible for one party to agree to an action independently). There are two layers of consensus: on the blockchain level for state replication and application level for an agreement lifecycle state machine. The system is implemented as a blockchain-based web application using Hyperledger Fabric as the back-end and Node.js as the application server.

\subsection{Application Server} \label{sec:appserver}
The Node.js application server (described interchangeably as Node or Node.js) acts as a REST server for the user interface and an integration point for multiple enterprise services such as Sharepoint, LDAP, email server, and document management. Primarily, the application server provides the application logic, which includes 
%includes user interface support
security mechanisms (including access control and authentication), process management, and blockchain transaction management. The main role of the application server in our performance tests was to provide the connection between the blockchain network (Hyperledger Fabric) and the application through the Hyperledger Fabric client SDK \cite{nodesdk}. The client SDK handles the blockchain transaction management and drives the transaction lifecycle outlined below. After the transaction is created in the application, the SDK opaquely handles interacting with the Hyperledger Fabric components (i.e., peers and orderers described below). As an enterprise application, there is support for user access control and authentication, logging (using log schemas), error handling, and transaction tracing. We emphasize that the application used in our testing runs in a production environment.

\subsection{Hyperledger Fabric} \label{sec:fabric}

\subsubsection{Fabric Overview.} Hyperledger Fabric is an open-source permissioned blockchain solution published under the Hyperledger Project \cite{hyperledgercommunity}. A Fabric network is composed of three main components: peers, orderers, and certificate authorities. Peers host the ledger and smart contracts (chaincode) and through a delivery service, are capable of delivering a range of events to clients (e.g., transaction statuses and successful commits). The current state of the blockchain is stored in a state database (stateDB) that the peer connects to (each peer has a state database). Additionally, each peer has a local ledger file that contains the consensus-driven chained blocks of executed transactions (i.e., the blockchain). A peer's role in the network can be endorser (i.e., execute transactions against the smart contract) and/or committer (i.e., all peers are committers). Orderers are responsible for implementing a deterministic consensus protocol that performs transaction ordering so that all peers execute the same transactions in the same sequence. An atomic broadcast API is exhibited by the orderers to guarantee total order of transactions. Lastly, certificate authorities are abstracted as membership services and assign identities to network participants, providing a permissioned network. Separate from the blockchain network is a client, which in our case is a Node.js application server,  that drives transactions to the network and receives events from network components. Hyperledger Fabric also provides the notion of channels that support confidentiality through data segregation.

\subsubsection{Transaction Flow.} 
\begin{figure}[t]
\captionsetup{font=footnotesize}
  \centering
      \includegraphics[width=0.9\textwidth]{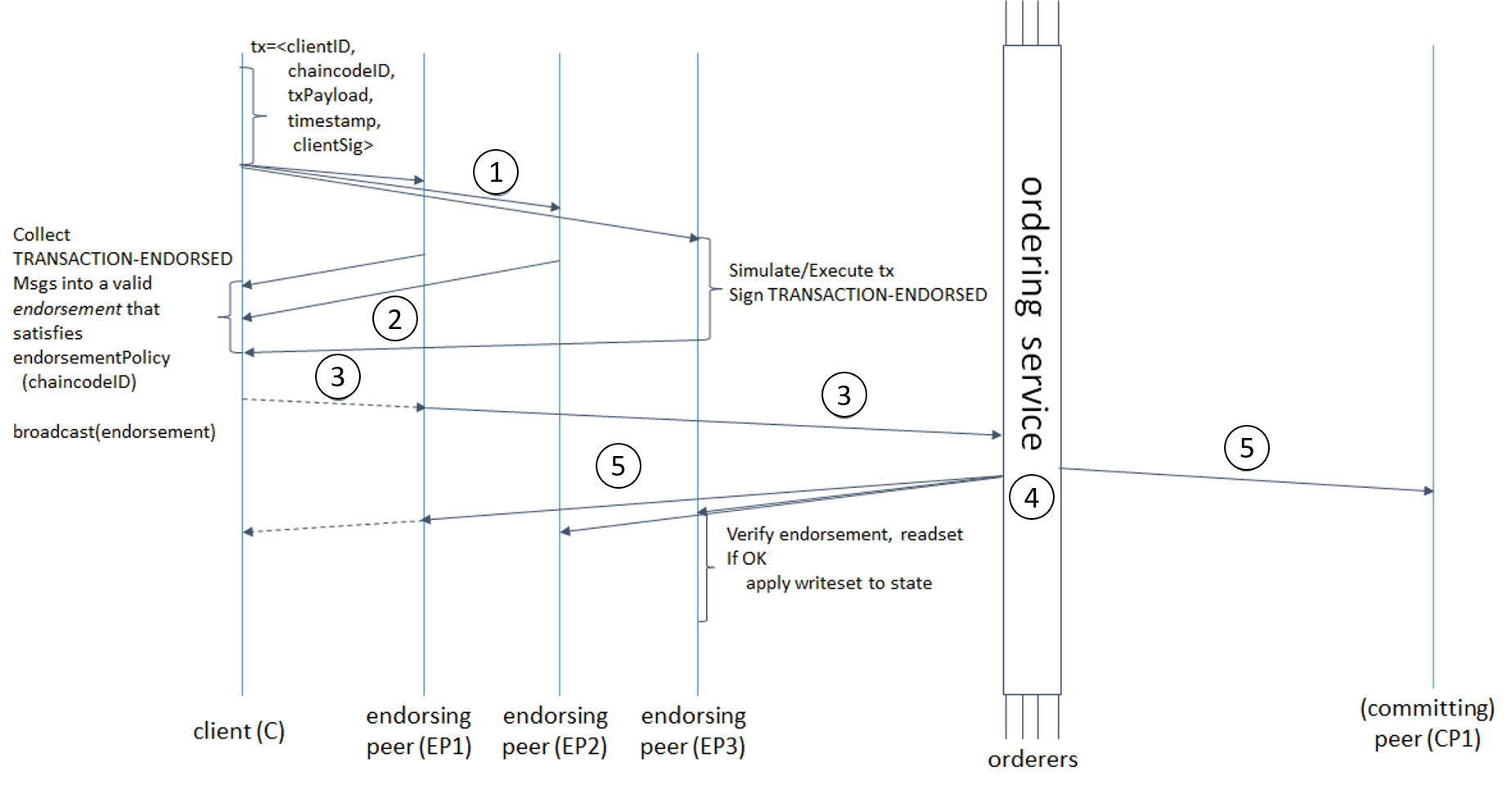}
  \caption{Hyperledger Fabric Transaction Flow \cite{txflow}}
   \label{fig:tx flow}
\end{figure}

There are three phases to transaction execution: (i) propose and endorse; (ii) order; (iii) validate and commit. Figure \ref{fig:tx flow} illustrates the transaction flow for the common case. 

\begin{enumerate}
	\item Clients first create, sign, and propose a transaction (containing chaincode name and parameters) to endorsing peers. The endorsing peers validate against an access control list (ACL) based on the client's signature and execute the transaction with their chaincode. Results are captured in the form of a Read-Write set (RWSet) and the peers provide attestation of execution result by signing the RWSet hash. This is just a simulated transaction execution, the ledger state is not updated at this point.
	\item If the transaction execution is successful, the endorsing peers send a transaction endorsement message back to the client.
	\item After receiving enough endorsements from peers (to match an endorsement policy), the client will compose a transaction from the endorsing responses and the initial transaction proposal. The transaction is signed by the client and sent to the ordering service (i.e., a cluster of orderers).
	\item The orderer deterministically orders the transactions it receives and creates a block of transactions. The block is then sent to all committing peers. 
	\item The committing peers validate the transactions in the received block and commit the transactions to their local ledger file (and update their state database). If the client is subscribed to the event service, they will be notified by peers when the transaction has been committed.
\end{enumerate}

%% file: Sections/Performance_Optimization_Methodology.tex
\section{Performance Optimization Methodology} \label{sec:methodology}

\begin{figure}[t]
\captionsetup{font=footnotesize}
  \centering
      \includegraphics[width=0.9\textwidth]{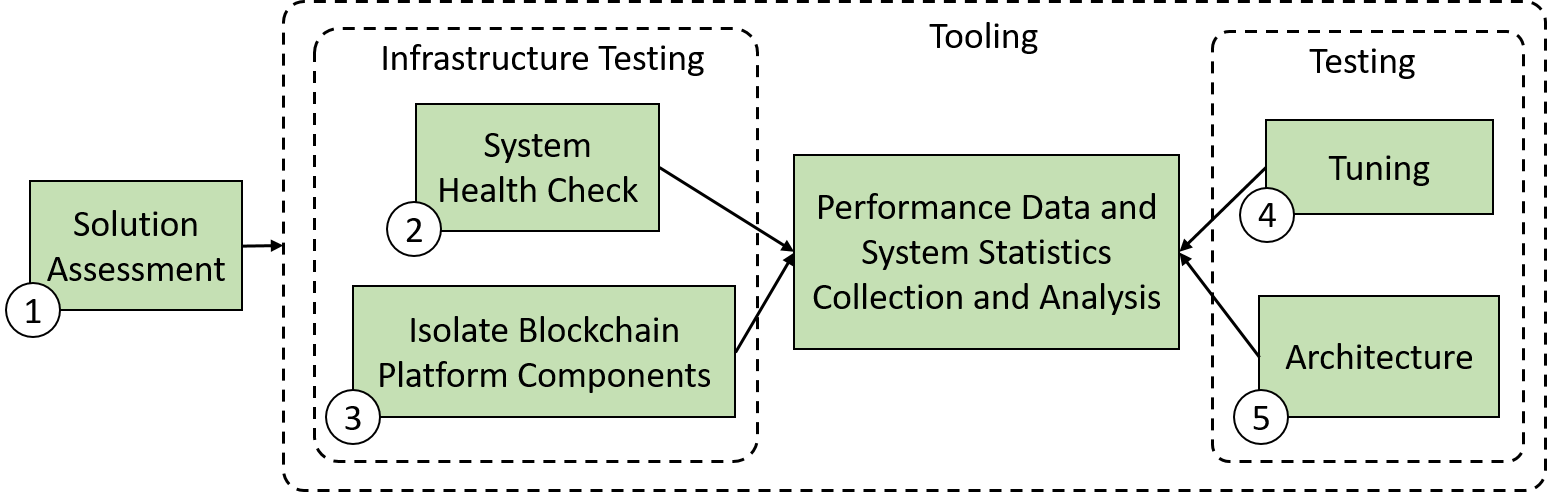}
  \caption{Performance Optimization Methodology}
   \label{fig:perf opt method}
\end{figure}

In this section, we present our performance optimization methodology, which is illustrated in Figure \ref{fig:perf opt method}. Leveraging this methodology helps in reducing the complexity of tuning a distributed system such as a blockchain application. Importantly, a methodology allows us to determine not just what or how to optimize, but \textit{why} the changes affect overall system performance. For example, isolating the application components to trace a transaction through its lifecycle and observing the causality of parameter tuning greatly improves the understanding of why performance improves or degrades. Although the methodology is general enough to be applied to a number of distributed systems, in this paper we focus on applying it to a blockchain-based system (specifically Hyperledger Fabric).

\subsection{Methodology Overview}

\subsubsection{Solution Assessment.}
The first step is to perform a solution assessment, which includes an architecture review and an analysis of the application implementation. As an initial step, the architecture review is meant to quickly determine if the application components or infrastructure need immediate attention (e.g., is there high resource contention between components? Are the components equally distributed across infrastructure layers?). Important aspects to examine in the application implementation are the data model (e.g., JSON object structure), smart contract implementation and complexity, REST gateways for API communication, and the end user interface. 

\subsubsection{System Health Check.}
The next step dives deeper into the infrastructure by performing a system health check. Before testing individual system components, we must check that the state of the system is acceptable for testing. This includes determining if the test environment has enough CPUs, memory, etc. Sections \ref{sec:solution assess} and \ref{sec:cloud} describes the on-premise (Figure \ref{fig:infra}) and cloud (Figure \ref{fig:cloudenv}) infrastructures the application is deployed on, respectively. 
% illustrates the infrastructure our application is deployed on.
In order to test aspects such as I/O and network, we leveraged the Flexible I/O Test (fio) \cite{fiodocs} and iPerf \cite{iPerf} tools, respectively.

\subsubsection{Isolate Blockchain Components.}
After tuning and testing the infrastructure, we can isolate individual components that run on this infrastructure. The main components of Hyperledger Fabric are the peers, the peers' stateDB (e.g., CouchDB), and the orderers. This step is meant to understand which phase of a transaction interacts with each component and where potential bottlenecks occur, allowing us to understand their internal limit when not throttled by the rest of the system. The main steps of the peers are proposal execution (endorsement) and transaction validation and commit. Breaking down these steps allows us to capture transaction proposal execution time, chaincode execution time, block validation time, and block commit time to the stateDB. On the stateDB level, we can examine the effect of batch updates to the database, the overall size of the database, and the duration of commits. For the orderer, we can look at the time spent gathering transactions and cutting blocks of transactions, the validation of the transactions in terms of structure and syntax, and the consensus protocol.

\subsubsection{Application \& Component Tuning.}
Now that the blockchain components have been isolated and analyzed, the fourth step is to tune the components through their parameters. For example, tuning block parameters, such as size and timeout on the blockchain layer, or configuring worker threads on the application server. There is also the challenge of optimizing these values for peek and regular traffic. The transaction validation process on the peers can be further parallelized with the validator pool size parameter to specify how many threads to spawn for validation. The total number of transactions in a block and the block cut timeout can be tuned in the orderer. For CouchDB, we can tune the underlying data structure and indexing strategy. Application tuning involves configuring the interactions with the Hyperledger Fabric network and the implementation of transaction management. Section \ref{sec:params} will provide a deeper analysis of the tunable system parameters and application code tuning.

\subsubsection{Architecture Tuning.}
For architectural tuning, we can look at vertical and horizontal scaling, caching data where appropriate, such as look up data in the smart contracts, and the distribution of components on the underlying infrastructure. In particular, horizontally and vertically scaling the architecture can have significant performance implications by adding more machines and components or more CPU and memory, respectively. In some cases, this could be the most expensive tuning option, which means it should be performed as the last step in the methodology.

\subsubsection{Statistics Collection \& Analysis.}
Throughout the methodology, tooling frameworks are used to assist with analysis by collecting performance data and statistics. Our performance analysis relied on four collection methods: (i) Prometheus \cite{prometheus} and Grafana \cite{grafana}; (ii) logs; (iii) tracing; and (iv) JMeter \cite{jmeter}. The main source of our blockchain statistics comes from Hyperledger Fabric exposing metrics, where Prometheus and Grafana consume and visualize the data, respectively. These metrics capture transaction flow phases and provide \textit{aggregated} data. This data allows us to determine if changes made to the system have positive or negative effects and the causal relations between components. For capturing \textit{specific} data points, we leverage logs from the application server and peers that provide client-side and system performance perspectives, respectively. 
%This data is similar to the statistics from the metrics, except these are \textit{specific} data points.
For example, we can record individual block validation times for a peer. We have a logging framework in place that collects all log records from application components for ease of data analysis.

Even with metrics and logs, it may be difficult to detect which component causes problems. Tracing can track a set of services participating in some task and their interactions. A trace is a directed acyclic graph that represents a hierarchy of spans for a request \cite{wattenhofer2019}. A span is a timed operation and can capture causal relations between spans. We leverage the OpenTracing framework \cite{opentracing} and Jaeger \cite{jaeger} for measuring API calls to the Hyperledger Fabric network. All transactions are traced based on their transaction ID and a unique request ID, which allows us to monitor specific transaction data. The tracing data is collected through our logging framework and transactions can be traced through the application components.

Finally, JMeter provides our reported throughput (i.e., transactions per second), the average latency of the requests, the total time of the test, and the error rate. The data reported from JMeter is typically the first data points we examine and based on those results, we analyze the aggregated data in Grafana and the specific data points captured in the logs and tracing.

%% file: Sections/Experiment_Setup.tex
\section{Experiment Environment \& Summary} \label{sec:solution assess}

%This section covers the solution assessment and system health check steps of the performance optimization methodology. 
This section covers the setup of our on-premise Z system experiments. We describe the infrastructure (Section \ref{sec:infrastructure}) and test environment (Section \ref{sec:test_env}), the main data model of our solution (Section \ref{sec:data_model}), and the application and smart contract implementation (Section \ref{sec:app_cc_imp}). A summary of our experiments with key findings is presented in Section  \ref{sec:summary}.

\subsection{Infrastructure} \label{sec:infrastructure}
\begin{figure}[t]
\captionsetup{font=footnotesize}
  \centering
      \includegraphics[width=0.9\textwidth]{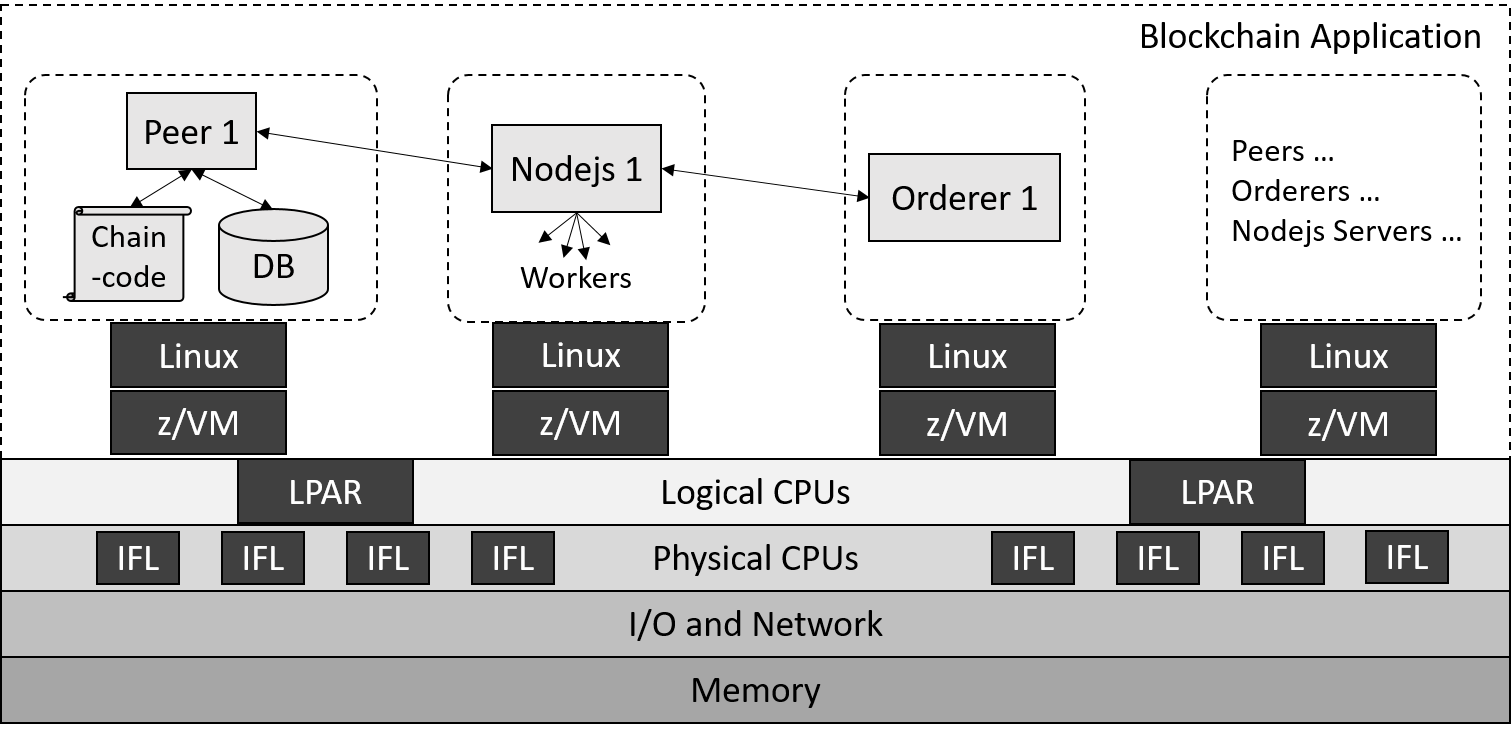}
  \caption{On-Premise Blockchain Application Infrastructure}
   \label{fig:infra}
\end{figure}

The blockchain-based application we performed our testing on is deployed to Red Hat Enterprise Linux (release 7.6 Maipo) running on z/VM (Figure \ref{fig:infra}). The z/VM operating system runs on the zSeries (z14) mainframe servers. Each architectural component of the application (or set of components) runs on one of these Linux VMs. Integrated Facility for Linux (IFLs), which is a dedicated Linux engine for exclusively processing Linux workloads \cite{zsystem}, are enabled in the infrastructure. Logical partitions (LPARs) provide the ability to share a single server among separate operating system images \cite{zsystem}. The IFL engines are assigned to LPARs and provide two layers of dynamic CPU mapping: physical IFL CPU cores to LPAR logical CPU cores and LPAR logical CPU cores to VM virtual CPU cores. In our environment, we have two LPARs at our disposal that support 5 to 13 VMs depending on the test configuration (the VMs are distributed across the LPARs). Each LPAR has up to 16 physical CPU cores (IFLs) since we tested with 4, 8, and 16 IFLs. All of the results reported in this paper are in the context of a 6:1 virtual to physical CPU mapping (Section \ref{sec:conlusion} discusses how this mapping can be improved to optimize CPU utilization). Refer to Figure \ref{fig:infra} for an illustration of the application infrastructure. This testing environment simulates the production environment the application runs in.

\subsection{Test Environment} \label{sec:test_env}
%{\color{red} Illustrate test setup (network, etc.). Unless otherwise specified, the results discussed below were determined from this test environment.}
Leveraging the above infrastructure, we deployed the application across VMs. The components of the Hyperledger Fabric blockchain platform were run in Docker containers (Docker Enterprise v17.06.2 and Docker Compose v1.20.0) and used version 1.4.1 images \cite{dockerHLF} (all containers use the standard open-source Fabric images provided by Hyperledger; there was no source code forking). The application servers are implemented with Node.js v8.10.0 and use the Hyperledger Fabric Node.js Client SDK version 1.4.0 to interact with the blockchain network. The chaincode transaction logic is implemented in Go (v1.10.4).

\begin{figure}[t]
\captionsetup{font=footnotesize}
  \centering
      \includegraphics[width=0.8\textwidth]{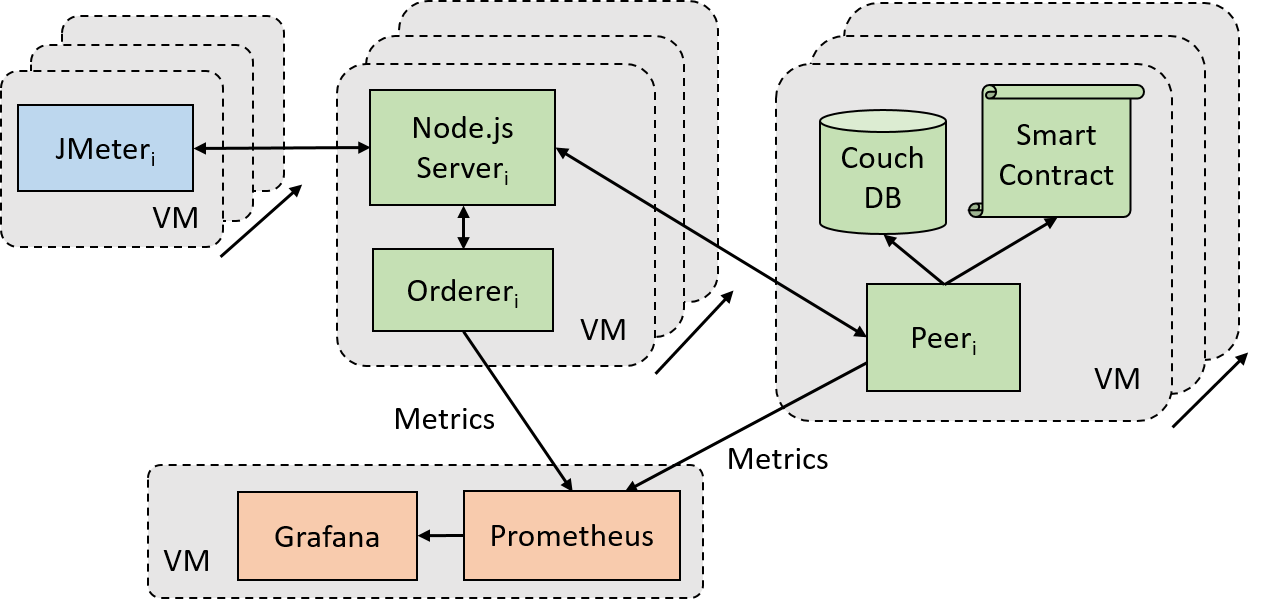}
  \caption{Blockchain Application Performance Test Topology}
   \label{fig:topology}
\end{figure}

Figure \ref{fig:topology} depicts the application topology used throughout our experiments. The performance experiments were run by multiple JMeter machines (v5.0 with Java 1.8) pushing REST requests to the application servers, which were connected to the Hyperledger Fabric blockchain network. The client load is simulated by these JMeter machines. The Node.js application servers connect to the peers and orderers of the blockchain network through the client SDK. All of the components in the test environment can be horizontally scaled out across VMs. Since the orderer containers consume minimal CPU cores ($< 10\%$) and to keep a minimal infrastructure imprint, the Node.js application servers and the orderer containers share the same VM. Metrics are exposed from the orderers and peers, which Prometheus (v2.6.1) consumes and is visualized by Grafana (v6.2.0). All experiments were carried out using production quality code and features (e.g., user access control mechanisms, data caching, transaction logging and tracing, error handling).

The Hyperledger Fabric network leveraged a single application channel composed of 2 organizations. A channel defines a network of organizations and their respective peer nodes \cite{channel}. The application's network setup is typically 2 peers per organization. 
%During our peer scaling experiments detailed below, the peers were scaled in one organization while the other organization had 2 peers. 
%Based on the application design, the first organization is considered the primary organization and selected for transaction endorsement, so scaling the peers in the second organization would not affect the endorsement procedure. 
An endorsement policy defines the set of organizations required to endorse a transaction for it to be considered valid. 
%Although the application uses one organization to endorse transactions, 
Our endorsement policy is set as any organization member can endorse a transaction (i.e., OR(org1.members, org2.members)) \cite{endorsepolicy}. Although the endorsement policy leverages an \textit{OR} clause, the application defines one organization as primary, and all transactions are endorsed by peers belonging to this organization.
%However, for the performance tests, we choose the first organization to endorse transactions.
Unless otherwise stated, the ordering service was composed of 3 Raft ordering nodes. 

\subsection{Data Model} \label{sec:data_model}

In Section \ref{sec:solution desc}, we discussed that the solution we conducted our performance testing on is modeled around the notion of an agreement. The core functionality of the system is to manage the lifecycle of the agreement data objects. An agreement data object is a JSON object composed of four parts: flat attributes and three arrays of embedded objects. Listing \ref{lst:agr object} outlines the agreement data model. Although the agreement object is implemented in JSON, we illustrate the attribute types here for completeness. Lines 2-7 are flat attributes composed of string, double, and date types, whereas lines 8-30 are arrays of embedded objects. Fundamentally, the agreement is between a local and foreign business unit, where all details of the agreement are formalized and approved in the data model. The data object can dynamically grow depending on the number of foreign business units. Additional agreement information is captured in the supporting documents on line 18, however, in our testing we omit any supporting documents. Lastly, an agreement can be in a number of states such as created, pending approval, and closed, which is captured in the statuses array. We have omitted some attributes from the listing. 

With the data model in Listing \ref{lst:agr object}, an average Hyperledger Fabric transaction size is about 7.5KB. The size can be broken down to the input parameters and the Read-Write set. The input parameters include the agreement data object, which is an average of 2730 bytes, and user metadata, which is 402 bytes. The Read-Write set contains the agreement object again (2730 bytes), the client certificate (800 bytes), and the peer certificate (800 bytes). There are some additional fields that contribute a few hundreds of bytes. Although the JSON data model comprises a large portion of the transaction size, this is a typical representation of business use cases where multiple attributes capture the necessary details for the business requirements.

\begin{lstlisting}[language=json,firstnumber=1,caption={Agreement Object Data Model},captionpos=b,label={lst:agr object}]
Agreement {
  _id: String
  type: String
  description: String
  creationDate: Date
  localBUName: String
  ...
  foreignBURechargeDetails: Array:Document
    foreignBURechargeDetail {
      foreignBUName: String
      foreignBURecharge: Array:Document
	foreignBURecharge: {
	  foreignBUAmount: Double
	  foreignBUApproverId: String			
	  ...		
	}
    }
  supportingDocuments: Array:Document
    supportingDocument {
      url: String
      attachmentDate: Date	
      ...		
    }
  statuses: Array:Document
    status {
     creationDate: Date
     description: String
     initiator: String 
     ...
    }
  ...
}

\end{lstlisting}

\subsection{Application \& Chaincode Implementation} \label{sec:app_cc_imp}
In a production deployment, the agreement object is created by the user through the UI and inserted to the blockchain network through the application server and chaincode. For our testing deployment, the agreement object is pre-populated and pushed to the application server from JMeter. User access control, error handling, transaction creation, and tracing initialization is handled by the application server. The chaincode manages the lifecycle of an agreement through a state machine and supports CRUD operations for agreement objects. User authentication, field level data validation, data caching, error handling, and transaction tracing is also supported in the chaincode.

\subsection{Summary of Experimental Results} \label{sec:summary}

Table \ref{tab:summary} shows a summary and rough performance improvement of the main experiments detailed in subsequent sections. The columns are organized based on the application and system components, the experiment parameter, the percentage of improvement performance (throughput), the configuration change that provided the improvement, and the section that provides detailed results and analysis. Due to the complex nature of distributed systems, these performance improvements are highly dependent on factors such as transaction load, infrastructure, application architecture, implementation, etc. The sections referenced in the table will provide a deeper description of the experiment results and how these results were achieved. Some configurations are not listed here, such as client configuration (Section \ref{sec:app_params}) and component deployment distribution (Section \ref{sec:rec_comp_dist}) since they are implicitly leveraged in many of the experiments.

\begin{table}
\captionsetup{font=footnotesize}
\centering
\resizebox{\textwidth}{!}{
\begin{tabular}{|>{\centering\arraybackslash}m{2cm}|>{\centering\arraybackslash}m{4cm}|>{\centering\arraybackslash}m{1.5cm}|>{\centering\arraybackslash}m{5cm}|>{\centering\arraybackslash}m{1.3cm}|}
  \hline
   \textbf{Comp.} & \textbf{Param.} & \textbf{Perf. Improv.} & \textbf{Config. Change} & \textbf{Sect.} \\ \hline
   \multirow{2}{*}{Peer}  & Validator Pool Size & 10\% & 16 to 100 pool size & \S\ref{sec:peer_params}\\ \cline{2-5}
   &  Endorsing Peer Scaling & 40\%  & 1 to 3 endorsers & \S\ref{sec:peer_params},\ref{sec:comp_scaling_peer} \\ \hline
    \multirow{3}{*}{StateDB} & B-tree Chunk Size (CouchDB) & 20\% & 256 to 4096 chunk size & \S\ref{sec:couch_params},\ref{sec:specialty_couch} \\ \cline{2-5}
    & Document ID (CouchDB) & 5\%  & Sequential IDs to Monotonic IDs based on timestamps & \S\ref{sec:couch_params},\ref{sec:specialty_couch} \\ \cline{2-5}
     &  StateDB Choice & 35\%  & CouchDB to GolevelDB & \S\ref{sec:app_params},\ref{sec:specialty_statedb}  \\ \hline
   \multirow{2}{*}{Orderer} &  Block Size & 7\%  & 250 to 500 transactions per block & \S\ref{sec:orderer_params},\ref{sec:specialty_bs}  \\ \cline{2-5}
   &  Consensus & 7\%  & Application sends endorsed txs to Raft leader & \S\ref{sec:orderer_params} \\ \hline 
   \multirow{3}{*}{App Server} &  Event Handling Strategy & 70\%  & Synchronous to asynchronous (fire-and-forget) commit confirmation  & \S\ref{sec:app_params}, \ref{sec:specialty_null}\\ \cline{2-5}
   &  Cluster & 44\%  & 12 to 32 workers & \S\ref{sec:specialty_null}  \\  \cline{2-5}
    %Client Config & 5\%  & x event source,x end peers & \ref{sec:app_params} \\ \hline
   &  Application Server Scaling & 25\%  & 2 to 6 servers & \S\ref{sec:app_params},\ref{sec:comp_scaling_app} \\ \hline % 2 to 6 servers
\end{tabular}}
\caption{Experiment Results Summary} 
\label{tab:summary}
\end{table}

%% file: Sections/Tunable_Components.tex
\section{Tunable Application Components} \label{sec:params}

In this section, we isolate the platform components and describe the tunable parameters that affect the performance of the system. Table \ref{tab:parameters} groups the parameters by system component and we describe and analyze each component below (Sections \ref{sec:peer_params}, \ref{sec:couch_params}, \ref{sec:orderer_params}, \ref{sec:app_params}, respectively). Sections \ref{sec:exp_baseline_summ}, \ref{sec:comp_scaling}, and \ref{sec:specialty} will leverage these parameters for result analysis and provide details regarding the high-level results outlined in this section.

\begin{table}
\captionsetup{font=footnotesize}
\centering
\resizebox{\textwidth}{!}{
\begin{tabular}{|c|c|c|c|}
  \hline
    \textbf{Peer (\S\ref{sec:peer_params})} & \textbf{CouchDB (\S\ref{sec:couch_params})} & \textbf{Orderer (\S\ref{sec:orderer_params})} & \textbf{App Server (\S\ref{sec:app_params})} \\ \hline
    Validator Pool Size & B+ Tree Chunk Size & Block Size & Event Handling Strategy\\ 
    Endorser & Monotonic ID & Block Cut Time & Event Hub \& Listeners \\ 
    Committer & Erlang VM Tuning & Consensus & Cluster \\
 &   &  & Configuration \\ \hline
\end{tabular}}
\caption{Tunable Parameters and Configurations for Application Components} 
\label{tab:parameters}
\end{table}

\subsection{Peer} \label{sec:peer_params}
\subsubsection{Validator Pool Size \& GOMAXPROCS.} In Hyperledger Fabric, the peer components are written in Go and are executed in the Go runtime environment. The Go runtime contains its own scheduler, apart from a kernel scheduler, that uses $m$:$n$ scheduling (i.e., $m$ goroutines are scheduled on $n$ OS threads) \cite{gobook}. 
%There is also a kernel scheduler that manages OS threads separately from the Go runtime scheduler, which manages goroutines. 
%A goroutine is a lightweight Go runtime thread. 
The GOMAXPROCS (GMP) parameter determines how many OS threads may be actively executing Go code simultaneously, which means GOMAXPROCS is the $n$ in $m$:$n$ scheduling \cite{gobook}.

The peer block validation routine is a computationally expensive task since all transactions in a block must be iterated through and transaction endorsements (e.g., digital signatures) are verified. To speed up block validation, the validator pool size (VPS) parameter sets an upper bound on the number of goroutines (lightweight runtime threads) that the peer will spawn during block validation for parallel processing. Validator pool size is implemented as a weighted semaphore and caps the number of concurrent validation goroutines and is the $m$ in $m$:$n$ scheduling. This means that there will be up to $m$ validation goroutines mapped across $n$ OS threads.

Since goroutines are cheaper (e.g., minimal/no context switching, variable stack size, go runtime scheduler) than OS threads, an ideal configuration is to use a low number of OS threads and a high number of goroutines. The exact number of OS threads and goroutines depends on the available CPU cores. By default, Hyperledger Fabric sets both VPS and GMP to the number of CPU cores available to the container. Based on our empirical results with 16 CPU cores, the optimized configuration of 100 VPS and 16 GMP provided a 10\% throughput boost over the default configuration. In Section \ref{sec:recommendations}, we describe how using a buffered channel for validation goroutine results may improve block validation performance.

\subsubsection{Endorser.} Peers can be assigned the endorser role to endorse transactions for an organization. If a peer is selected to be an endorser, they are responsible for executing transactions (i.e., the chaincode transaction logic is executed) and endorsing transactions by digitally signing them. In terms of CPU usage, the endorsing peers consume the most CPU cores out of the Hyperledger Fabric components. Therefore it is important to consider how their resource consumption will affect the underlying infrastructure and how many endorsers should be deployed on the network. 

\subsubsection{Committer.} All peers in a Hyperledger Fabric network commit transactions. However, peers that are not set as endorsers will only perform endorsement \textit{verification} and transaction commits. This is an important distinction since purely committing peers will consume less resources because they do not need to execute chaincode and endorse transactions. Since committing peers perform less processing, having the client application wait for transaction commit confirmations from committing peers can typically reduce latency compared to waiting for commit confirmations from endorsing peers (this may not be the case if, for example, committing peers have less resources available than endorsing peers, message delays, geographical location, etc.).

\subsection{CouchDB} \label{sec:couch_params}
\subsubsection{B+ Tree Chunk Size.} CouchDB uses an append-only B+ tree data structure to index documents and views, and to store the main database file  \cite{anderson2010}. Separate B-trees are used for the database and view index. 
%The B-tree implementation in CouchDB is append-only and supports Multi-Version Cocurrency Control (MVCC). 
For every document update, CouchDB will load the B-tree nodes from disk that point to the updated documents or, in the case of an insertion, the key range where the new document's ID would be located \cite{anderson2010b}. Typically, the B-tree nodes will be loaded from the filesystem cache, however, updates to documents in regions of the tree that have not been accessed in a while cause disk seeks. These disk seeks can block writing, which can affect other aspects of the system. Preventing these disk seeks can improve the overall performance of CouchDB \cite{anderson2010b}.

The order of a tree is the maximum number of pointers to subtrees from one node (i.e., the maximum number of elements in a node). In CouchDB, the order of the B-tree is determined by a chunk size \cite{dionne2012}. This chunk size can be modified through the $btree\_chunk\_size$ parameter, which is used in the $chunkify$ routine \cite{couchDBGit}. Changing this parameter provides a memory vs. speed trade-off \cite{newson2013}; higher values will use more memory and improve the speed of updates to the B-tree, whereas lower values will use less memory at the cost of update speed. Since the chunk size determines how large the nodes are, it has a direct correlation with the number of disk seeks required on an update operation. A larger chunk size results in less tree re-structuring since nodes will be filled up at a lower rate.

The default B-tree chunk size is 1279. From our experiments, increasing the chunk size to 4096 gave a 20\% TPS improvement. Higher values than 4096 did not improve performance (performance started to degrade with values over 6,000). Small chunk sizes, such as 256, also degrade performance.
%It is important to note that if the chunk size is set too small, than there will be a significant decrease in performance. For example, a chunk size of 256 caused a TPS degradation of 13\%.

\subsubsection{Monotonic ID.} %As mentioned above, CouchDB uses a B+ tree to store the main database file. 
 Documents are indexed in the CouchDB B-tree by their IDs. Prior to compaction, the choice of ID has a significant impact on the layout of the B-tree \cite{couchDocs}. Using monotonic IDs will minimize the number of intermediate tree nodes that need to be rewritten \cite{couchSeq}. Alternatively, random IDs cause intermediate nodes to be frequently rewritten, which results in decreased throughput and wasted disk space (because of the append-only structure of the B-tree) \cite{couchDocs}. The choice of ID also affects the caching behaviour since IDs clustered together will have more cache hits, which results in faster insertion time. Random IDs result in documents being inserted in arbitrary  locations in the B-tree, which may result in many cache misses.

Our document ID generation algorithm creates IDs based on the current timestamp when the document was created. Depending on the speed of request processing, timestamp-based IDs may result in ID conflicts if documents are generated at the same time. However, if two documents are inserted to CouchDB with the same ID, the system chaincode will handle this concurrency error by ensuring the key is first read before updating it (i.e., verifying Read-Write sets). 
%by accepting one document and rejecting the other since the document ID has a unique field constraint.  
%An ID clash would require a GET call to the stateDB and read set (contains a list of unique keys and their committed version numbers \cite{readset}) validation to resolve this conflict, which would slow down the overall system. 
To avoid ID collisions 
% between processes 
with high probability, the timestamps should be created with high precision. Additionally, IDs do not have to be consecutive, rather they just need to be ordered \cite{couchSeq}. Therefore, the timestamp ID generation follows these requirements. Using timestamps as IDs guarantees monotonicity since timestamps $TS_1 < ...  < TS_n$ are sequentially ordered. Using this timestamp based monotonic ID generation improved TPS by 5\%.

\subsubsection{Erlang VM Tuning.} CouchDB is implemented in Erlang, which runs in an Erlang VM (BEAM). Runtime-specific parameters can be configured through VM arguments and may improve CouchDB's performance. We evaluated a subset of parameters related to CouchDB performance.

By default, Hyperledger Fabric enables the $+K$ (kernel polling) and $+A$ (async thread pool) parameters, which can improve performance if there are many file descriptors in use and improve I/O operations, respectively. 
%The $+A$ parameter sets the number of threads in the async thread pool \cite{elrDocs}, where the default value is the number of cpu cores. 
%Since this parameter relates to I/O operations rather than cpu bound operations, and based on CouchDB's use in Hyperledger Fabric,
The default value of the async thread pool is the number of CPU cores and since there are more CPU bound operations than I/O bound, increasing this value had no positive effect. %Other than the default parameters, the remaining parameters did not positively impacted performance. 
%In fact, 
Changing the additional parameters from their default values degraded performance. These parameters included $+spp$ (port parallelism), $+stbt$ (scheduler binding), $ts$ (bounds scheduler threads across hardware threads), $ps$ (spreads schedulers across physical chips), $db$ (schedulers spread over processors), $+scl$ (scheduler load compaction),  $+sfwi$ (scheduler wakeup interval), and $+zdbbl$ (distribution buffer busy limit).

\subsection{Orderer} \label{sec:orderer_params}
\subsubsection{Block Size.}  The block size determines how many transactions the orderer will collect before cutting a block. A larger block size will result in more transactions to be validated and longer commit times. However, larger blocks will result in less blocks, which means there are less blocks to commit. From our experiments, an ideal block size for a specific load has a block fill ratio close to 100\% in order to reduce block cut timeouts. However, the total number of blocks is also important to consider with the block fill ratio, as too many blocks will have a negative impact on the commit procedure, especially for high transaction arrival rates. Due to the state database lock mechanisms and REST interface (CouchDB), block commits are a bottleneck in the transaction flow and minimizing the time spent performing this operation is ideal to maximizing performance. Therefore, an ideal block size configuration takes into account the number of blocks created and the fill ratio. Additionally, tuning CouchDB affects the commit times, so properly configuring the stateDB is also important.

\subsubsection{Block Cut Time (Batch Timeout).} The block cut time is a fallback mechanism if the block is not filled in a specific time. This value provides an upper bound for how long it takes for the block to be cut. Minimizing the delta between the time it takes to fill a block and the default cut time is important to reduce idle time for the orderer waiting for the cut timeout to occur. However, setting this value too low can potentially reduce the block fill ratio (since it could take longer than the timeout to fill the block). Under high load tests, the block cut time has the most impact during the beginning and end of the test (i.e., when the test is ramping up or down, the transaction rate is not high enough to fill the block). An optimal value should take into account the flow of traffic (i.e., periods of low or high load) to balance idle time waiting for blocks to be cut during low load and batching enough transactions into a block during peak load.
%and efficiently handle average work loads 

\subsubsection{Consensus.} 
%{\color{red}Raft vs Kafka. Node pointing to Raft leader.} 
The consensus protocol is responsible for deterministic ordering of transactions in blocks. As an integral component of a blockchain network the choice of protocol has an impact on performance. Hyperledger Fabric currently supports Kafka\footnote[1]{Kafka is on the road to being deprecated \cite{deprecateKafka}} \cite{kafka} and Raft-based \cite{raft} consensus. Raft is the recommended ordering service since it performs similar to Kafka while being easier to maintain \cite{ordering}, so we use Raft for the majority of the paper. Raft follows a leader-follower model, where the leader drives the ordering of transactions and replicates messages to the follower nodes. Leaders are chosen through an election campaign after followers have not received heartbeat messages in a set timeout. Since the leader will only be changed if the current leader node crashes (Raft is crash fault tolerant, not Byzantine fault tolerant), we can directly connect the application to the leader Raft node. Assuming Raft nodes are collocated in a local network, connecting directly to the leader improves performance, whereas connecting to followers redirects connections to the leader. In v1.4.1, Hyperledger Fabric selects the orderer Raft node to start an election campaign by checking which orderer ID equals $hash(channelID) \% clusterSize + 1$. On network startup, the same orderer Raft node will be elected the channel leader with high probability (assuming the same orderer nodes were present before). Therefore, we know which orderer will be elected the leader \textit{a priori}, so we can configure the application servers to directly connect to the leader. Of course, if the leader crashes, the application will have to be redirected to the new Raft leader.% (e.g., through a load balancer).

\subsection{Node.js Application Server} \label{sec:app_params}
\subsubsection{Event Handling Strategy.} The application server leverages an event handling strategy to determine how the client should wait for commit events emitted from peers after a transaction is committed to the ledger. The fabric-network SDK (v1.4.0 \cite{fabricnetwork}) provides five strategies: \texttt{MSP\_SCOPE\_ALLFORTX}, \texttt{MSP\_SCOPE\_ANYFORTX}, \texttt{NETWORK\_SCOPE\_ALLFORTX}, \texttt{NETWORK\_SCOPE\_ANYFORTX}, and \texttt{null}. The strategies differ in their \textit{scope} (msp or network) and \textit{policy} (all or any). The scope and policy refers to if the client needs to listen for transaction commit events from all or any peers on the network or organization level.
%Setting the event handler strategy to \texttt{MSP\_SCOPE\_ALLFORTX} means that the client will listen for transaction commit events from \textit{all} currently connected peers in the client identity's organization \cite{handlerStrategy}. 
%\texttt{MSP\_SCOPE\_ANYFORTX} is similar to \texttt{MSP\_SCOPE\_ALLFORTX} except the client will wait for successful events from \textit{any} peer.
%\texttt{NETWORK\_SCOPE\_ALLFORTX} and \texttt{NETWORK\_SCOPE\_ANYFORTX} are similar to the previous strategies but have a broader scope (i.e., events from peers in the network rather than just on the organization level). 
The \texttt{null} strategy means transaction invocations return immediately to the client after the endorsed transaction is successfully sent to the orderer (i.e., fire and forget) \cite{handlerStrategy}. Transactions are still eventually committed on all peers, however the client does not wait for a commit confirmation. The choice of strategy has a great effect on the overall system performance, where in terms of decreasing performance: \texttt{null} $>>$ \texttt{NETWORK\_SCOPE\_ANYFORTX} $\geq$ \texttt{MSP\_SCOPE\_ANYFORTX} $>$ \texttt{MSP\_SCOPE\_ALLFORTX} $>$ \texttt{NETWORK\_SCOPE\_ALLFORTX}. 

The network scope strategies require either all peers (potentially slow based on the network size) or a single (any) peer 
%(i.e., fastest peer) 
to confirm the transaction commit. For small networks (e.g., two organizations) a single peer commit confirmation is fast, but may become slower with larger networks due to the number of network connections between the client and peers. The organization scopes' (MSP) performance is bounded by the network scopes since only a subset of peers are available in the MSP  strategies.

The \texttt{null} strategy results in much greater performance than the previous strategies because the client does not need to wait for transactions to be committed before finishing the request. The transaction commit is the slowest operation in the transaction flow, so the bottleneck is removed by this strategy (requests are not bounded by the commit operation). However, this performance increase comes at the cost of acknowledging failed transaction commits, which is not acceptable in some use cases. The \texttt{null} strategy is beneficial to determine an upper bound in performance to compare with the other commit strategies and to help tune transaction commit related processes.

Compared to a \texttt{null} commit strategy, \texttt{network any} and \texttt{all} strategies give 40\% and 50\% throughput degradation, respectively (with a four peer network). The MSP based strategies have similar degradation of 40\% to 45\%. All peers in the network perform commit operations and process the same block, thus consume CPU cycles. Although all peers process blocks, only the \texttt{all} policy commit strategies will be negatively impacted by the number of peers in the network.

\subsubsection{Event Hub \& Listeners.} In order to support an asynchronous design, applications should register a listener to be notified of events \cite{nodesdk}. Committing peers provide an event stream to publish events to registered listeners \cite{nodesdk}. Event hubs and listeners are related to the event handling strategy since they are the mechanisms in which the client is notified of events and how the strategy is satisfied. An event hub resides on the application server (provided by the Fabric SDK) and manages the events emitted from peers. Listeners can be registered in the event hub to listen for blocks, transactions (which leverage block events), and chaincode events. An event hub needs to be registered and a listener established in order to begin the monitoring of events. The choice of event hub and listener can play a large role in performance.

We began our testing with the peer level event hub, which was present in v1.1 to v1.2 of the Fabric Node.js SDK. A peer level event hub resulted in multiple timeouts due to being tied to the peer, which  severely degraded performance. The Node.js SDK v1.3 improved the event hub by tying it to the channel rather than the peer. Tying the event hub to the channel provided more stability and improved the performance compared to the peer event hub. Upgrading to the v1.4.0 SDK provided the best performance with the channel-based event hub and abstracting away all event hub and listener setup to a single SDK method call (\textit{transaction.submit} method \cite{txsubmit}).

\subsubsection{Cluster.} A Node.js application runs in a single thread and does not leverage multiple cores. In order to leverage multi-core systems, Node.js provides the option to launch a cluster of Node.js processes. When the cluster is enabled, there will be a master process that is responsible for distributing incoming connections in a round-robin fashion to the worker processes. Typically, each worker process is bound to a CPU core (i.e., usually the optimal number of workers is the number of CPU cores available). More workers will allow for more concurrency and greatly improve throughput and latency compared to a single instance of a Node.js application with no clustering enabled. However, each worker will have duplicate block processing since every worker receives block events from the network (i.e., a worker notifies the client of a transaction commit by processing block events). Fundamental to the Node.js architecture, the event loop may be blocked due to the computational complexity of JavaScript callbacks \cite{nodejsEventLoop}. Since block event processing is completed through callbacks, the event loop is periodically blocked since there are multiple workers processing the block. Although this contributes to increasing latency, the benefit of leveraging clustering outweighs the impact of event loop blocking.

\subsubsection{Client Configuration.} The application layer of a Hyperledger Fabric network can specify the roles of the peers in the connection profile configuration \cite{networkconfig}. Two important peer roles are \textit{endorsing peer} and \textit{event source}. If a peer is set to an endorsing peer then they will execute and endorse transactions. Alternatively, non-endorsing peers will just commit transactions. Setting a peer as an event source means that the application will only accept events (such as commit confirmations) from peers listed as an event source. An optimal configuration is to separate the endorsing peers from the event source peers. If committing peers (i.e., not endorsing) are set as event sources, then they will typically respond back to the application with commit confirmation events faster than endorsing peers since they have less processing to do.

%% file: Sections/Experiment_Summary.tex
\section{Experiment Baseline \& System Health Check} \label{sec:exp_baseline_summ}

In this section, 
%we take the analysis of the platform components from the previous section and apply them to our performance testing. 
we begin with a description of the baseline performance test (Section \ref{sec:exp_baseline}) and provide a summary of our system health check activities (Section \ref{sec:exp_summ}) for preliminary tuning to get the application to an acceptable testing state.  This section provides the starting point for the performance experiments reported in subsequent sections. Unless otherwise stated, all performance tests are with data insertion workloads (query results are discussed in Section \ref{sec:query}). All reported throughput results in this paper are only calculated based on transactions with a success status; no experiments had failed transactions.

\subsection{Baseline} \label{sec:exp_baseline}
Our initial blockchain application performance benchmarks reported 30 TPS and 6s latency. This result was captured from 5 JMeter machines (with 30 threads and 100 loops each), Hyperledger Fabric v1.1 with a 150 transaction block size, 3 Kafka-based orderers, and 4 peers (1 endorser), 2 Node.js application servers, and 4 CPU cores per VM (there were 5 VMs in the network). This is the baseline configuration from which we apply our performance optimization methodology.

\subsection{System Health Activity Summary} \label{sec:exp_summ}

We first reviewed the current state of the application, network topology, and baseline performance testing results in order to identify and address possible bottlenecks. By reviewing the application implementation, we observed that client, channel, event hub, and user context objects were not being reused across requests and the configuration data was loaded and parsed for each request (i.e., expensive I/O operations). Additionally, the transaction event hub disconnected after every request, which affected other request processing by the application server because of the repeated connection closing. The combination of these connection issues resulted in thousands of open gRPC connections recreated for each request. Modifying the application implementation to cache and reuse the client, channel, and event hub connections improved throughput by 100\% to 60 TPS.

We scaled the application servers from 2 to 4 servers and added an additional endorsing peer to divide the transaction workload to 2 individual peers (since the single endorsing peer's CPU was saturated). Detailed results and analysis of application and peer scaling is reported in Section \ref{sec:comp_scaling}. The combination of the client connection reuse, removal of event hub disconnection, reuse of gRPC connections, endorsement peer load balancing, and application server instance scaling resulted in 90 TPS.

Since the application leveraged an early version of Hyperledger Fabric (v1.1), the blockhchain platform was upgraded to v1.3. Version 1.3 of Hyperledger Fabric significantly improved performance since some of the locks within Hyperledger Fabric were removed \cite{HLF13Changes}. Most notably, the event hub model was redesigned to provide a more reliable and efficient block delivery service. This redesign was the main contributor to a 44\% throughput improvement to 130 TPS.

We began horizontally scaling the blockchain components across VMs (increasing the infrastructure footprint from 5 to 7 VMs) to avoid resource contention and vertically scaled the CPU cores on each VM to 8 cores. The better load distribution across the underlying infrastructure and the increased availability of resources increased the throughput to 310 TPS. The testing environment was composed of 6 JMeter machines (each with 350 threads), 3 Node.js application servers (each with 12 workers), 400 transaction block size, and 3 endorsing peers.

Finally, we upgraded Hyperledger Fabric to the latest version at the time of writing (v1.4.1) \cite{HLF14}. Fabric v1.4 provides an improved programming model that adds a layer of abstraction to the client SDK. The Kafka-based consensus protocol used in previous versions of Fabric was replaced with a Raft-based ordering service, which is easier to maintain and deploy as it is built in to the peer process. The improved Fabric platform can process transactions at a higher rate, so we applied preliminary block size, application server worker, and load tuning to increase throughput to 600 TPS. 
%{\color{red}Fabric 1.4 upgrade (new programming model, replaced Kafka with Raft consensus). block size tuning, workers, load. 550 to 620 TPS.} All results in the remainder of the paper leverage Hyperledger Fabric v1.4.1.
With the initial bottlenecks discovered and addressed, as well as preliminary performance test results, we provide an in-depth analysis in the subsequent sections.

%% file: Sections/Component_Scaling.tex
\section{Component Scaling Results \& Analysis} \label{sec:comp_scaling}
This section provides the detailed results and analysis of our component scaling experiments and is organized by application server (Section \ref{sec:comp_scaling_app}) and peer scaling (Section \ref{sec:comp_scaling_peer}).

\subsection{Application Server Scaling} \label{sec:comp_scaling_app}
A blockchain-based application involves many CPU-bound operations that originate from the application servers and the Hyperledger Fabric components. For example, the application servers manage multiple open connections and transaction listeners while the Fabric components perform operations such as transaction endorsement, validation, and commit, which include computations such as digital signature creation and verification. In order to effectively accommodate these operations, we vertically scaled the infrastructure to 16 CPU cores per virtual machine. Unless otherwise stated, the remainder of the test results are based on VMs with 16 CPU cores.

Table \ref{tab:server scaling} shows the results of horizontally scaling the application servers. The test cases are grouped based on the level of concurrent requests coming in to the test environment (i.e., base, mid, high). The \texttt{base} class (TC1 to 4) refers to a low number of threads submitting transactions from JMeter (e.g., 300 to 600 threads per JMeter). The \texttt{mid} (TC5 to 7) and \texttt{high} (TC8 to 10) classes spawn more threads to drive the transactions to the network (e.g., 600 to 900 and 1200 to 1600 threads per JMeter, respectively). Based on our test results, the optimal peer configuration is to map one endorsing peer to one Node server (see peer scaling below). Therefore, all of the following tests have a separate endorsing peer per Node server. In a production environment, it is necessary to take into account high availability, which means the application server needs a 1:$n$ application server and peer mapping (a single peer still endorses transactions, but there are several additional peers for resilience). For the practicality of performance testing, we deploy a strict 1:1 mapping.

\begin{figure}[t]
  \centering
  %\hspace{-1cm}
  \begin{subfigure}[b]{0.5\textwidth}
    \captionsetup{font=footnotesize}
\begin{tikzpicture}
     \begin{axis}[
     scale=0.6,
       xmin = 0, xmax = 11,
       ymin = 550, ymax = 900,
       axis y line*=left,
       xlabel={Test Case (TC)},
       xlabel near ticks,
       ylabel={Throughput (TPS)},
       ylabel near ticks,
       xtick={1,2,3,4,5,6,7,8,9,10},
       xticklabels={1,2,3,4,5,6,7,8,9,10},
        label style={font=\tiny},
  tick label style={font=\tiny},
    grid = major,
  grid style = {dashed, gray!30},
legend pos=south east,
legend style={nodes={scale=0.5, transform shape}},
legend columns=1,
     ]
       \addplot[
    color=red,
    mark=square,
    mark size=1pt,
    ]
    coordinates {
(1,686)(2,736)(3,837)(4,855)
    }; \label{p1}
    \addplot[
    color=blue,
    mark=square,
    mark size=1pt,
    ]
    coordinates {
(5,890)(6,892)(7,889)
    }; \label{p2}
        \addplot[
    color=green,
    mark=square,
    mark size=1pt,
    ]
    coordinates {
(8,801)(9,841)(10,871)
    }; \label{p3}
     \end{axis}
     \begin{axis}[
     scale = 0.6,
       xmin = 0, xmax = 11,
       ymin = 1, ymax = 12,
       hide x axis,
       axis y line*=right,
       ylabel={Avg Latency (s)},
       ylabel near ticks,
        xtick={1,2,3,4,5,6,7,8,9,10},
       xticklabels={1,2,3,4,5,6,7,8,9,10},
        label style={font=\tiny},
  tick label style={font=\tiny},
  legend pos=south east,
legend style={nodes={scale=0.4, transform shape}},
legend columns=1,
     ]
     \addlegendimage{/pgfplots/refstyle=p1}\addlegendentry{Base (TPS)}
       \addlegendimage{/pgfplots/refstyle=p2}\addlegendentry{Mid (TPS)}
         \addlegendimage{/pgfplots/refstyle=p3}\addlegendentry{High (TPS)}
        \addplot[
    color=red,
    mark=*,dashed,
    mark size=1pt,
    ]
    coordinates {
(1,2.5)(2,3.2)(3,6.1)(4,11.6)
    }; \addlegendentry{Base (Lat)}
            \addplot[
    color=blue,
    mark=*,dashed,
    mark size=1pt,
    ]
    coordinates {
(5,4.2)(6,6.4)(7,8.7)
    }; \addlegendentry{Mid (Lat)}
                \addplot[
    color=green,
    mark=*,dashed,
    mark size=1pt,
    ]
    coordinates {
(8,8.8)(9,11.2)(10,11.7)
    }; \addlegendentry{High (Lat)}
     \end{axis}
   \end{tikzpicture}
\caption{Throughput and Latency}\label{fig:throughput_lat_app_scale}
\end{subfigure}
  %\hspace{0.5cm}
  %\hfill
  \begin{subfigure}[b]{0.4\textwidth}
    \captionsetup{font=footnotesize}
\begin{tikzpicture}
\begin{axis}[
  %width=\columnwidth, height=6cm,  % size of the image
  scale = 0.6,
  grid = major,
  grid style = {dashed, gray!30},
legend pos=north east,
legend style={nodes={scale=0.5, transform shape}},
legend columns=1,
ylabel near ticks,
  % xmode=log,log basis x=10,
  % ymode=log,log basis y=10,
  xmin = 0,   % start the diagram at this x-coordinate
  xmax = 60,  % end   the diagram at this x-coordinate
  ymin = 0,   % start the diagram at this y-coordinate
  ymax = 100, % end   the diagram at this y-coordinate
  %/pgfplots/xtick = {0,5,...,60},  % make steps of length 5
  xtick={0,10,20,30,40,50,60},
  xticklabels={$t_0$,$t_1$,$t_2$,$t_3$,$t_4$,$t_5$,$t_6$},
  axis background/.style = {fill=white},
  ylabel = {CPU Usage (\%)},
  %ylabel shift = -2pt,
  xlabel = {Time Steps},
   label style={font=\tiny},
  tick label style={font=\tiny},]
  
  \addplot[color=blue,mark=o,line width=0.5pt,mark size=1pt] table[col sep=comma]{Data/2nodes_ulzkvkd9_nodejs_cpu.csv};
  \addplot[color=red,mark=o,line width=0.5pt,mark size=1pt] table[col sep=comma]{Data/3nodes_ulzbcd11_nodejs_cpu.csv};
\addplot[color=green,mark=o,line width=0.5pt,mark size=1pt] table[col sep=comma]{Data/4nodes_ulzkvd10_nodejs_cpu.csv};
\legend{TC5, TC6, TC7}
\end{axis} 
\end{tikzpicture}
\caption{CPU Usage for App Servers}\label{fig:cpu usage app serv}
  \end{subfigure}
  \caption{Impact of App Server Scaling on Throughput/Latency and CPU}
\end{figure}
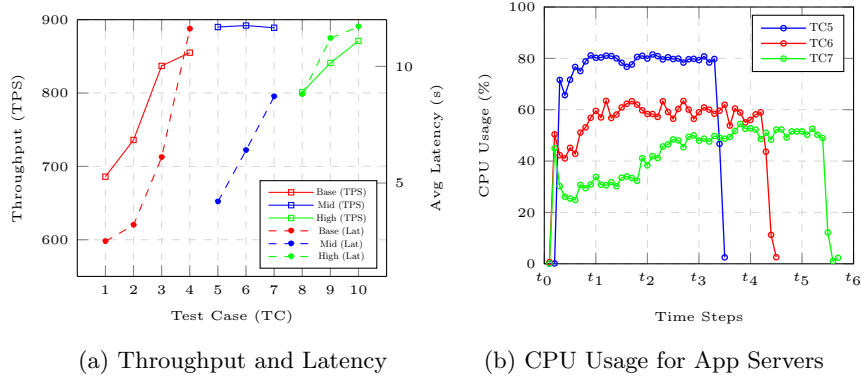

To avoid CPU contention, the Node servers and endorsing peers run on VMs that reside on different LPARs. Since we have two LPARs at our disposal, we must balance the peers and Node servers across VMs on these LPARs. For example, in the 2 Node test, there is one endorsing peer and Node server running on LPAR 1 and the same number of components running on LPAR 2. The Node servers have the highest CPU utilization followed by the endorsing peers, so evenly distributing these components across the underlying infrastructure is crucial. Figure \ref{fig:cpu usage app serv} depicts the CPU usage for the Node.js application servers 
%and endorsing and committing peers 
for the \texttt{mid} test classes in Table \ref{tab:server scaling}. 

\begin{table}[t]
\captionsetup{font=scriptsize}
\centering
\resizebox{\textwidth}{!}{
\begin{tabular}{|c||c|c|c||c|c||c|c|c||c|c|c|}
  \hline
  \multirow{10}{*}{} 
      & \multicolumn{3}{c||}{\textbf{JMeter}}   
          & \multicolumn{2}{|c||}{\textbf{Node.js Server}}            % \cline{2-10}
   		 & \multicolumn{3}{|c||}{\textbf{Hyperledger Fabric}}            % \cline{2-9}
			& \multicolumn{3}{|c|}{\textbf{Results}} \\             \cline{2-12}
  & \textbf{Clts} & \textbf{Thr.} & \textbf{Lps} & \textbf{Srvs} & \textbf{Wrks} & \textbf{BS} & \textbf{TO (s)} & \textbf{End.} & \textbf{Thrpt} & \textbf{Avg. Lat. (ms)} & \textbf{BFR}\\  \hline
  \textbf{TC1 (Base)} & 6 & 300 & 100 & 2 & 12 & 400 & 2 & 2 & 686 & 2557 & 97\%  \\      \hline
\textbf{TC2 (Base)} & 6 & 400 & 100 & 3 & 24 & 400 & 3 & 3 & 736 & 3243 & 98\%  \\      \hline
\textbf{TC3 (Base)} & 8 & 600 & 100 & 4 & 24 & 600 & 3 & 4 & 837 & 6066 & 98\%  \\      \hline
\textbf{TC4 (Base)} & 6 & 1600 & 25 & 6 & 24 & 800 & 3 & 6 & 855 & 11561 & 92\%  \\      \hline
  \textbf{TC5 (Mid)} & 6 & 600 & 100 & 2 & 24 & 600 & 2 & 2 & 890 & 4221 & 97\% \\      \hline
\textbf{TC6 (Mid)} & 6 & 900 & 100 & 3 & 30 & 800 & 3 & 3 & 892 & 6406 & 92\% \\      \hline
 \textbf{TC7 (Mid)} & 8 & 900 & 100 & 4 & 32 & 700 & 3 & 4 & 889 & 8701 & 96\% \\      \hline
\textbf{TC8 (High)} & 6 & 1200 & 25 & 2 & 24 & 1200 & 4 & 2 & 801 & 8760 & 94\% \\      \hline
\textbf{TC9 (High)} & 6 & 1600 & 25 & 3 & 24 & 1600 & 4 & 3 & 841 & 11201 & 93\% \\      \hline
\textbf{TC10 (High)} & 8 & 1200 & 45 & 4 & 32 & 1200 & 3 & 4 & 871 & 11665 & 83\% \\      \hline

\end{tabular}}
\caption{Application Server Scaling Results. Table columns are organized by components and test results. For JMeter, we report the number of instances (Clts), threads (Thr.), and loops (Lps). We report the total number of Node.js application servers (Srvs) and worker threads (Wrks); threads are evenly distributed across servers. Hyperledger Fabric parameters are the number of transactions in a block, or block size (BS), block cut timeout (TO), and the number of endorser peers (End.). The results are reported as throughput (Thrpt), average latency (Avg Lat), and the block fill ratio (BFR). Table rows are organized based on the test classification (base, mid, high) and case (TC).} 
\label{tab:server scaling}
\end{table}

Comparing the base case results for 2, 3, 4, and 6 application servers shows that applying horizontal scaling to the application servers improves throughput (Figure \ref{fig:throughput_lat_app_scale} depicts the throughput and latency for each test case of Table \ref{tab:server scaling}). Starting  with 2 Node servers gave us a baseline of 686 TPS. With 6 worker threads per Node server, the CPU utilization of the Node servers was between 40\% and 45\%.  This shows that there is enough resources to increase the number of worker threads per node, which will improve throughput and latency. For the next test with 3 Node servers, we increased the workers to 24 in order to utilize more CPU cores, which resulted in the Nodes' CPU usage increasing to 75\%. We also increased the number of JMeter threads to 400 since we now have an additional application server to handle requests. These changes directly impacted the throughput, which increased to 736 TPS. The average transaction latency increased by 1 second since there are more concurrent transactions being executed and processed. The next two base cases (TC4, 5) resulted in 837 TPS and 855 TPS by scaling the application servers to 4 and 6, respectively. The increase in application servers and the number of endorsing peers allowing for more concurrent transaction processing is the main contributing factor for the throughput increase. Six application servers resulted in high contention for CPU cores, so in order to get a comparable throughput as previous tests, the threads were increased to 1600 (i.e., there needs to be a large number of concurrent requests for the overall throughput to increase).

For the mid cases, we increased the load to 600 threads per JMeter for the 2 application server configuration. With a previous block size of 400 transactions, the orderer was able to cut the block in an average of 0.56 seconds, which is well below the 2 second block cut timeout. Since the block cut time was very low, the block fill ratio was 97\%, which fits in an optimal range for the number of blocks propagating through the system. Increasing the block size while keeping the load constant will result in a lower block fill ratio, which means more blocks will reach the 2 second cut timeout and lower the overall throughput and latency (due to the timeout). Matching the block size to the load is necessary to keep the optimal block fill ratio, which may be difficult in an enterprise setting where application volumes will fluctuate. Increasing the concurrent load caused the throughput to jump to 890 TPS. Since the number of incoming requests is increasing, the average response latency increases as well. 
%Figure \ref{fig:cpu usage app serv} shows that the committing and endorsing peers' CPU usage is constant at 15\% and 25\%, respectively. 
The application server CPU usage is much greater than the peers, where a topology with 2 application servers utilizes 80\% of the CPU cores. As the number of application servers increases, the average CPU usage for one server instance decreases because there is contention between servers for the shared physical cores of the underlying infrastructure. This also verifies that the application server is the most computationally intensive component in a Hyperledger Fabric network for data insertion loads (Section \ref{sec:specialty} shows the throughput and latency improvement when maximizing the application servers CPU usage with the null commit strategy).
The remainder of the \texttt{mid} test cases show that the infrastructure resources are being exhausted 
% (mainly CPU cores) 
since there is no improvement to throughput and latency is further increasing. 

The high load test cases further verify that there are not enough resources to support over 1200 threads across 6 JMeters. With such a large number of concurrent users, the block size must also be large to prevent too many block commits (since committing the block is a slow operation). However, the large block size also means that the time to fill the block will increase, which affects the latency (this ``queuing'' can be seen in the rightmost points of Figure \ref{fig:throughput_lat_app_scale}). The combination of high concurrent requests increasing the application server CPU usage and the time it takes to cut and commit the block contributed to the throughput decreasing from the \texttt{mid} test case results.

\subsection{Peer Scaling} \label{sec:comp_scaling_peer}
Scaling the application servers allows more transactions to be concurrently sent to the Hyperledger Fabric network. However, without also properly scaling the number of peers endorsing transactions, the benefit of additional application servers is lost due to the bottleneck in the peer processing. The results of the endorsing peer scaling tests are in Table \ref{tab:peer scaling}. 

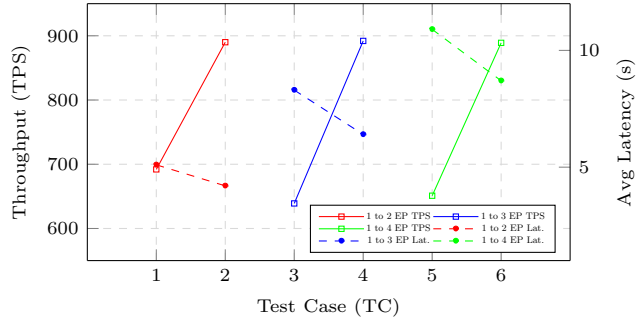
\begin{figure}[t]
  \centering
    \captionsetup{font=footnotesize}
\begin{tikzpicture}
     \begin{axis}[
     width=8cm, height=5cm,  % size of the image
       xmin = 0, xmax = 7,
       ymin = 550, ymax = 950,
       axis y line*=left,
       xlabel={Test Case (TC)},
       xlabel near ticks,
       ylabel={Throughput (TPS)},
       ylabel near ticks,
       xtick={1,2,3,4,5,6},
       xticklabels={1,2,3,4,5,6},
        label style={font=\scriptsize},
  tick label style={font=\scriptsize},
    grid = major,
  grid style = {dashed, gray!30},
legend pos=south east,
legend style={nodes={scale=0.5, transform shape}},
legend columns=2,
     ]
       \addplot[
    color=red,
    mark=square,
    mark size=1pt,
    ]
    coordinates {
(1,692)(2,890)
    }; \label{pp1}
    \addplot[
    color=blue,
    mark=square,
    mark size=1pt,
    ]
    coordinates {
(3,639)(4,892)
    }; \label{pp2}
        \addplot[
    color=green,
    mark=square,
    mark size=1pt,
    ]
    coordinates {
(5,651)(6,889)
    }; \label{pp3}
     \end{axis}
     \begin{axis}[
     width=8cm, height=5cm,  % size of the image
       xmin = 0, xmax = 7,
       ymin = 1, ymax = 12,
       hide x axis,
       axis y line*=right,
       ylabel={Avg Latency (s)},
       ylabel near ticks,
        xtick={1,2,3,4,5,6},
       xticklabels={1,2,3,4,5,6},
        label style={font=\scriptsize},
  tick label style={font=\scriptsize},
  legend pos=south east,
legend style={nodes={scale=0.4, transform shape}},
legend columns=2,
     ]
     \addlegendimage{/pgfplots/refstyle=pp1}\addlegendentry{1 to 2 EP TPS}
       \addlegendimage{/pgfplots/refstyle=pp2}\addlegendentry{1 to 3 EP TPS}
         \addlegendimage{/pgfplots/refstyle=pp3}\addlegendentry{1 to 4 EP TPS}
        \addplot[
    color=red,
    mark=*,dashed,
    mark size=1pt,
    ]
    coordinates {
(1,5.1)(2,4.2)
    };  \addlegendentry{1 to 2 EP Lat.}
            \addplot[
    color=blue,
    mark=*,dashed,
    mark size=1pt,
    ]
    coordinates {
(3,8.3)(4,6.4)
    }; \addlegendentry{1 to 3 EP Lat.}
                \addplot[
    color=green,
    mark=*,dashed,
    mark size=1pt,
    ]
    coordinates {
(5,10.9)(6,8.7)
    };  \addlegendentry{1 to 4 EP Lat.}
     \end{axis}
   \end{tikzpicture}
  \caption{Impact of Endorser Peer (EP) Scaling on Throughput/Latency}\label{fig:throughput_lat_peer_scale}
\end{figure}

The first two test cases (TC1 and TC2) in Table \ref{tab:peer scaling} compare the mid-level 2 Node server results from Table \ref{tab:server scaling} with the same configuration. Test case 1 reduces the number of endorsing peers to 1, whereas TC2 has 2 endorsing peers. 
%with only one endorsing peer. 
As can be seen from the results of TC1 and TC2, there is a 29\% increase in throughput when the number of endorsing peers matches the number of application servers (692 TPS vs. 890 TPS)\footnote[2]{The impact of scaling the number of endorsing peers is also discussed in \cite{ferrispeers}}. The throughput and latency results for each test case of Table \ref{tab:peer scaling} are shown in Figure \ref{fig:throughput_lat_peer_scale}.

\begin{table}
\captionsetup{font=footnotesize}
\centering
\resizebox{\textwidth}{!}{
\begin{tabular}{|c||c|c|c||c|c||c|c|c|c||c|c|c|}
  \hline
  \multirow{6}{*}{} 
      & \multicolumn{3}{c||}{\textbf{JMeter}}   
          & \multicolumn{2}{|c||}{\textbf{Node.js Server}}            % \cline{2-10}
   		 & \multicolumn{4}{|c||}{\textbf{Hyperledger Fabric}}            % \cline{2-9}
			& \multicolumn{3}{|c|}{\textbf{Results}} \\             \cline{2-13}
  & \textbf{Clts} & \textbf{Thr.} & \textbf{Lps} & \textbf{Srvs} & \textbf{Wrks} & \textbf{BS} & \textbf{TO (s)} & \textbf{End.} & \textbf{\#Peers}& \textbf{Thrpt} & \textbf{Avg. Lat. (ms)} & \textbf{BFR}\\  \hline
  \textbf{TC1} & 6 & 600 & 100 & 2 & 24 & 600 & 2 & 1 & 2 & 692 & 5137 & 98\% \\      \hline 
  \textbf{TC2} & 6 & 600 & 100 & 2 & 24 & 600 & 2 & 2 & 2 & 890 & 4221 & 97\% \\      \hline
   \textbf{TC3} & 6 & 900 & 100 & 3 & 30 & 800 & 3 & 1 & 6 &639 & 8347 & 98\% \\      \hline
   \textbf{TC4} & 6 & 900 & 100 & 3 & 30 & 800 & 3 & 3 & 6 &892 & 6406 & 92\% \\      \hline
 \textbf{TC5} & 8 & 900 & 100 & 4 & 32 & 700 & 3 & 1 & 8 & 651 & 10957 & 98\% \\      \hline
  \textbf{TC6} & 8 & 900 & 100 & 4 & 32 & 700 & 3 & 4 & 8 & 889 & 8701 & 96\% \\      \hline
\end{tabular}}
\caption{Endorsing Peer Scaling Results} 
\label{tab:peer scaling}
\end{table}

The endorsement of transactions by peers is a main step in the Hyperledger Fabric transaction flow. 
%A main step in the transaction flow of a Hyperledger Fabric network is peers endorsing transactions. 
If there is only one peer given the endorser role, then all transactions sent to the network will funnel through this peer (there could be many peers in the network, but all non-endorser peers will be committers). For example, in TC1 of Table \ref{tab:peer scaling}, there are 2 Node.js application servers handling transactions. However, with only one peer endorsing transactions, that peer is a bottleneck in the system. By assigning a second peer as an endorser and mapping the endorsers to the application servers (i.e., 1:1 mapping), we increase the parallelization of the system since each peer will handle their respective application server's transactions. Latency is also improved because of the distributed load across peers (i.e., one peer is not overloaded). The same pattern applies to test cases 3 to 6 (1:1 mapping of application servers to endorsing peers improves both throughput and latency). Test case 5 yields the highest latency (11s) since there are 4 application servers sending traffic to 1 endorsing peer, causing a bottleneck in the transaction flow.

\begin{table}
\captionsetup{font=scriptsize}
\centering
\resizebox{\textwidth}{!}{
\begin{tabular}{|P{1.5cm}||P{1cm}||P{1.5cm}|P{1.5cm}|P{1.5cm}||P{1cm}|P{1cm}|P{1cm}|P{1.5cm}|P{1.5cm}|P{1cm}||P{1cm}||P{1cm}|}
  \hline
  \multirow{4}{*}{}
      & \multicolumn{1}{c||}{\textbf{Endorse}}   
          & \multicolumn{3}{|c||}{\textbf{Block Creation}}            % \cline{2-10}
   		 & \multicolumn{6}{|c||}{\textbf{Validate \& Commit}}            % \cline{2-9}
			& \multicolumn{1}{|c||}{\textbf{Raft}} 
                      & \multicolumn{1}{|c|}{\textbf{JMeter}} \\             \cline{2-13}
  & \textbf{Avg time to complete proposal (ms)}  & \textbf{Avg block cut time (ms)} & \textbf{Avg time to validate tx (ms)} & \textbf{Avg time to enqueue tx (ms)} & \textbf{Avg time to validate block (ms)} & \textbf{Avg time for ledger block processing (ms)} & \textbf{Avg time to commit block to storage (ms)} & \textbf{Avg time to commit block changes to statedb (ms)} & \textbf{CouchDB processing time - BatchUpdateDocs (ms)} & \textbf{Avg time to commit block (ms)} & \textbf{Raft data persist duration (ms)} & \textbf{Avg latency (ms)}\\  \hline
  \textbf{TC1: 2N, 1E} & 296 & 981 & 1 & 1.1 & 106 & 87 & 62 & 430 & 122 & 728  & 8.7 & 5100  \\      \hline
  \textbf{TC2: 2N, 2E} & 294 & 870 & 1 & 1.2 & 115 & 92 & 68 & 473  & 139 & 798 & 9.5 & 4200  \\      \hline
\textbf{TC5: 4N, 1E} & 672 & 1084 & 0.7 & 1.7 & 135 & 110 & 83 & 707  & 168 & 1099 & 9.6 & 10900  \\      \hline
\textbf{TC6: 4N, 4E} & 274 & 1137 & 0.9 & 0.8 & 138 & 106 & 78 & 576  & 185 & 951 & 9.8 & 8700  \\      \hline
\end{tabular}}
\caption{Endorser Impact Results. 
The table is organized based on the three main phases of the Hyperledger Fabric transaction flow: transaction endorsement (peer), block creation (orderer), and transaction validation \& block commit (peer). For peer transaction endorsement, we capture the \textit{average time the endorsing peers take to complete a proposed transaction} from the client (includes chaincode execution). On the orderer, we capture the \textit{average time to cut a block} (i.e., block generation), \textit{average time to validate a transaction} (time spent receiving the transaction message from the client, unmarshaling message contents, validating the client's signature, and readying the message to be enqueued), and \textit{average time to enqueue a transaction} (time spent enqueuing, i.e. ordering, a transaction through the Raft consensus protocol). 
Committing peer block validation and commit includes the \textit{average time to validate a block} (verifying endorsement signatures on all transactions in the block), 
\textit{average time for ledger block processing} (validating the state and read/write sets for all transactions in the block), 
\textit{average time to commit the block to storage} (adding a commit hash composed of block metadata to the block and committing the block to the local ledger file),
\textit{average time for committing block changes to stateDB} (updating the world state in the stateDB based on transactions in the block), \textit{CouchDB processing time} from the BatchUpdateDocs function (the time take for BatchUpdateDocs to complete the request to CouchDB; this function is called when updating the stateDB), and 
\textit{average time to commit a block} (summation of previous 4 metrics). The final columns are \textit{Raft data persist duration} (time for a Raft node to store it's entries, state, and snapshot) and the \textit{average latency} reported from JMeter (entire transaction lifecycle, including the application server processing). Apart from JMeter latency, this data was gathered through the Hyperledger Fabric metrics service. 
} 
\label{tab:endorser impact}
\end{table}

Table \ref{tab:endorser impact} shows the transaction lifecycle breakdown for the 2 and 4 endorser peers in Table \ref{tab:peer scaling} (TC1, TC2, TC5, TC6). Comparing TC1 and TC2 (rows 1 and 2, respectively) shows that having 2 application servers with 1 endorsing peer each improves the block cut time by 13\%, but slightly increases the block commit time by an average of 70ms. With only 1 endorsing peer, all transactions sent from the application servers funnel through this single peer, which creates a bottleneck in the system and lowers the transaction rate to the orderer. The lower transaction rate increases the time it takes for the block to be filled. An additional endorser allows transactions from each application server to be endorsed in parallel, which improves the rate of transactions being sent to the orderer and decreases the time to fill a block. This also affects the block commit rate since blocks are being generated quickly and the peers are committing blocks more often (which results in the increased block commit time). The higher block generation rate decreases latency since the application does not need to wait as long for the transactions to complete and is the main reason for the 29\% throughput improvement.% (692 TPS vs. 890 TPS).

Rows 3 and 4 report the transaction breakdown for 4 application servers with varying numbers of endorsers. Compared to the 2 Node.js server results, the difference in proposal execution between 1 and 4 endorsers is much greater with 4 servers. This is due to the further parallelization of endorsement with 2 more endorsing peers than in rows 1 and 2. However, 4 servers with 4 endorsing peers has differing effects on block commits than 2 servers with 2 endorsers; the former decreases the commit time (1099ms to 951ms) whereas the latter increases the commit time (728ms to 798ms). The number of clients and the amount of requests increased in TC5 and TC6 (8 clients and 900 threads per client), as well as a block size of 700 transactions, which caused the time to cut the block to increase (i.e., more transactions are required to fill the block). This positively affects the average block commit time because the block fill ratio is lower than the 1 endorser test (98\% for 1 endorser, 96\% for 4 endorsers). 
With more blocks containing less than 700 transactions (96\%), the average block commit time is faster because of the higher number of smaller blocks. 
%With more blocks that are under the 700 block size, on average, the block commit is faster because of the higher number of smaller blocks. 
This illustrates the balance between how many blocks are created, the ratio of filled blocks, and the number of transactions in the block. A fill ratio difference of 2\% saved an average of 150ms to the block commit process. However, as the block fill ratio decreases, the throughput will be negatively affected since the block cut timeout will be reached too frequently (i.e., the application must wait for the block to be cut).

%% file: Sections/Specialty_Tests.tex
\section{Specialty Tests Results \& Analysis}  \label{sec:specialty}
This section reports the specialty case tests that are not related to horizontal component scaling. These results include commit strategies, CouchDB tuning, and state database choice (e.g., CouchDB, GolevelDB). A detailed description of the parameters used in the following tests are provided in Section \ref{sec:params}.

\subsection{Null Commit Strategy}  \label{sec:specialty_null}
\begin{figure}[t]
  \centering
    \captionsetup{font=footnotesize}
\begin{tikzpicture}
     \begin{axis}[
     width=9cm, height=5cm,  % size of the image
       xmin = 0, xmax = 9,
       ymin = 550, ymax = 1900,
       axis y line*=left,
       xlabel={Test Case (TC)},
       xlabel near ticks,
       ylabel={Throughput (TPS)},
       ylabel near ticks,
       xtick={1,2,3,4,5,6,7,8},
       xticklabels={1,2,3,4,5,6,7,8},
        label style={font=\scriptsize},
  tick label style={font=\scriptsize},
    grid = major,
  grid style = {dashed, gray!30},
legend style={nodes={scale=0.4, transform shape},at={(0.5,0.03)},anchor=south},
legend columns=2,
     ]
       \addplot[
    color=red,
    mark=square,
    mark size=1pt,
    ]
    coordinates {
(1,600)(2,723)(3,857)(4,919)(5,971)(6,1005)(7,1039)(8,1863)
    }; \label{ppp1}
     \end{axis}
     \begin{axis}[
     width=9cm, height=5cm,  % size of the image
       xmin = 0, xmax = 9,
       ymin = 4, ymax = 16,
       hide x axis,
       axis y line*=right,
       ylabel={Avg Latency (s)},
       ylabel near ticks,
        xtick={1,2,3,4,5,6,7,8},
       xticklabels={1,2,3,4,5,6,7,8},
        label style={font=\scriptsize},
  tick label style={font=\scriptsize},
legend style={nodes={scale=0.4, transform shape},at={(0.5,0.03)},anchor=south},
legend columns=2,
     ]
     \addlegendimage{/pgfplots/refstyle=pp1}\addlegendentry{Throughput}
        \addplot[
    color=blue,
    mark=*,dashed,
    mark size=1pt,
    ]
    coordinates {
(1,15.5)(2,12.8)(3,10.7)(4,10)(5,9.5)(6,9.1)(7,8.8)(8,5)
    };  \addlegendentry{Latency}
     \end{axis}
   \end{tikzpicture}
  \caption{Impact of Null Commit Strategy on Throughput/Latency}\label{fig:throughput_lat_null}
\end{figure}
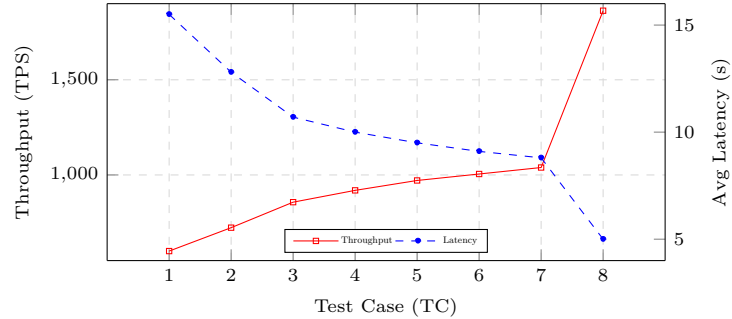

In order to determine if the selection of the commit strategy was a bottleneck in the system, we removed the need for transaction commit waiting on the client side by specifying the \textit{null commit strategy}. Table \ref{tab:null commit} shows the results of using the null commit strategy with 2 and 4 Nodes (all tests were run with 16 CPU cores). Test case 1 (TC1) provides a baseline test with the \texttt{NETWORK\_SCOPE\_ANYFORTX} strategy, which resulted in 600 TPS and 15.5s latency. Comparing TC1 to TC7 (the same configuration with null commit) shows a 73\% throughput increase to 1039 TPS and a reduction in average latency by 7 seconds. Scaling the application servers in TC8 further improves the throughput to 1863 TPS with 5s latency and demonstrates that transaction execution and endorsement on the peers can efficiently process large transaction loads (over 670,000 insertions) and the impact commit waiting has on the system. Figure \ref{fig:throughput_lat_null} illustrates the throughput and latency of the test cases in Table \ref{tab:null commit}.

The removal of the client waiting for commits affects multiple aspects of the system (TC2 to TC8). Since the application servers ``fire and forget" transactions, there is no need for event hubs and transaction listeners to monitor commit events. This frees up both CPU and memory that the application servers can leverage. The CPU usage of a single application server for TC3 to 7 is shown in Figure \ref{fig:cpu usage null commit}. The constraint that the listeners have on CPU resources is evident since, as we increment the number of worker processes, the Node application servers' CPU usage increases from 50\% to 100\% utilization. Now that the application servers are not waiting for transaction commit confirmations,
%and not fully utilizing available CPU cores, the application servers
 transaction processing in the application servers finishes quickly at maximum CPU utilization. The application servers can process more transactions, which fills the blocks faster and results in a 100\% block fill ratio. 
Regardless of the transaction processing rate on the application servers, the peers maintain constant CPU utilization (around 35\%) and finish processing (i.e., committing transactions) after the application servers complete their processing. The peers' constant CPU utilization can be attributed to the exclusive locking of the stateDB when committing transactions since this concurrency control limits the peers' transaction commit rate.

\begin{table}
\captionsetup{font=footnotesize}
\centering
\resizebox{\textwidth}{!}{
\begin{tabular}{|c||c|c|c||c|c|c||c|c|c||c|c|c|}
  \hline
  \multirow{8}{*}{} 
      & \multicolumn{3}{c||}{\textbf{JMeter}}   
          & \multicolumn{3}{|c||}{\textbf{Node.js Server}}            % \cline{2-10}
   		 & \multicolumn{3}{|c||}{\textbf{Hyperledger Fabric}}            % \cline{2-9}
			& \multicolumn{3}{|c|}{\textbf{Results}} \\             \cline{2-13}
  & \textbf{Clts} & \textbf{Thr.} & \textbf{Lps} & \textbf{Srvs} & \textbf{Wrks} & \textbf{Tx Wait} & \textbf{BS} & \textbf{TO (s)} & \textbf{End.} & \textbf{Thrpt} & \textbf{Avg. Lat. (ms)} & \textbf{BFR}\\  \hline
   \textbf{TC1} & 6 & 1600 & 25 & 2 & 32 & \cmark & 1600 & 4 & 2 & 600 & 15539 & 93\%  \\      \hline
  \textbf{TC2} & 6 & 1600 & 25 & 2 & 12 & \xmark & 1600 & 4 & 2 & 723 & 12844 & 100\%  \\      \hline
  \textbf{TC3} & 6 & 1600 & 25 & 2 & 16 & \xmark & 1600 & 4 & 2 & 857 & 10734 & 100\%  \\      \hline
  \textbf{TC4} & 6 & 1600 & 25 & 2 & 20 & \xmark & 1600 & 4 & 2 & 919 & 10024 & 100\%  \\      \hline
  \textbf{TC5} & 6 & 1600 & 25 & 2 & 24 & \xmark & 1600 & 4 & 2 & 971 & 9472 & 100\%  \\      \hline
 \textbf{TC6} & 6 & 1600 & 25 & 2 & 28 & \xmark & 1600 & 4 & 2 & 1005 & 9082 & 100\%  \\      \hline
 \textbf{TC7} & 6 & 1600 & 25 & 2 & 32 & \xmark & 1600 & 4 & 2 & 1039 & 8807 & 100\%  \\      \hline
 \textbf{TC8} & 8 & 1200 & 70 & 4 & 64 & \xmark & 1200 & 3 & 4 & 1863 & 5031 & 100\%  \\      \hline
\end{tabular}}
\caption{Null Commit Strategy with 2 and 4 Node Results} 
\label{tab:null commit}
\end{table}

Figure \ref{fig:commit null commit} illustrates the commit strategy's effect on the end-to-end transaction lifecycle, including transaction creation in the application server and transaction processing in Hyperledger Fabric (the reported duration does not include additional application tier and client processing). The data is reported from a 3 Node.js server test with 6 worker threads each, 800 block size, 1600 threads, 3 endorsers, and 16 CPU cores per VM. With a commit confirmation strategy, the duration of transaction processing linearly increases over time because of the transaction listeners consuming resources and the bottleneck of CouchDB interactions during block commits. Alternatively, the null commit strategy results in constant transaction processing duration because the bottleneck of block commits and transaction listener resource consumption are eliminated.

Omitting transaction commit confirmations is not practical for most applications since we want to make sure a transaction is committed before responding to the client. However, performing this test gives insight into how transaction listeners and transaction commits affect the overall system. This provides an upper bound on the system and a scale unit for the application servers. 
%These results provide the delta between null commit strategy and full commit. 
In Section \ref{sec:recommendations} we describe how an asynchronous request handling design can bridge the gap between a null commit strategy and full commit confirmation.

\begin{figure}[!tbp]
  \centering
  \hspace{-1cm}
  \begin{minipage}[b]{0.5\textwidth}
    \captionsetup{font=footnotesize}
\begin{tikzpicture}
\begin{axis}[
  %width=\columnwidth, height=7cm,  % size of the image
   scale=0.7,
  grid = major,
  grid style = {dashed, gray!30},
legend pos=north east,
legend style={nodes={scale=0.4, transform shape}},
legend columns=1,
ylabel near ticks,
  % xmode=log,log basis x=10,
  % ymode=log,log basis y=10,
  xmin = 0,   % start the diagram at this x-coordinate
  xmax = 40,  % end   the diagram at this x-coordinate
  ymin = 0,   % start the diagram at this y-coordinate
  ymax = 100, % end   the diagram at this y-coordinate
  %/pgfplots/xtick = {0,5,...,60},  % make steps of length 5
  xtick={0,5,10,15,20,25,30,35,40,45,50},
  xticklabels={$t_0$,$t_1$,$t_2$,$t_3$,$t_4$,$t_5$,$t_6$,$t_7$,$t_8$},
  axis background/.style = {fill=white},
  ylabel = {CPU Usage (\%)},
  xlabel = {Time Steps},
  tick align = outside,
  label style={font=\tiny},
  tick label style={font=\tiny},]
  
  \addplot[color=blue,mark=o,line width=0.5pt,mark size=1pt] table[col sep=comma]{Data/Null_Commit_CPU/ulzkvd10_857TPS.csv};
  \addplot[color=red,mark=o,line width=0.5pt,mark size=1pt] table[col sep=comma]{Data/Null_Commit_CPU/ulzkvd10_node_919tps.csv};
  \addplot[color=green,mark=o,line width=0.5pt,mark size=1pt] table[col sep=comma]{Data/Null_Commit_CPU/ulzkvd10_node_971tps.csv};
  \addplot[color=orange,mark=o,line width=0.5pt,mark size=1pt] table[col sep=comma]{Data/Null_Commit_CPU/ulzkvd10_node_1005tps.csv};
  \addplot[color=black,mark=o,line width=0.5pt,mark size=1pt] table[col sep=comma]{Data/Null_Commit_CPU/ulzkvd10_node_1039.csv};
\legend{857TPS, 919TPS, 971TPS, 1005TPS, 1039TPS}
\end{axis} 
\end{tikzpicture}
\caption{CPU Usage for Application Servers (AS) for Null Commit Strategy}\label{fig:cpu usage null commit}
  \end{minipage}
  %\hspace{1cm}
  %\hfill
  \begin{minipage}[b]{0.4\textwidth}
    \captionsetup{font=footnotesize}
\begin{tikzpicture}
\begin{axis}[
  %width=\columnwidth, height=6cm,  % size of the image
  scale=0.7,
  grid = major,
  grid style = {dashed, gray!30},
legend pos=north west,
legend style={nodes={scale=0.4, transform shape}},
legend columns=1,
ylabel near ticks,
  % xmode=log,log basis x=10,
  % ymode=log,log basis y=10,
  xmin = 0,   % start the diagram at this x-coordinate
  xmax = 20,  % end   the diagram at this x-coordinate
  ymin = 0,   % start the diagram at this y-coordinate
  ymax = 8, % end   the diagram at this y-coordinate
  %/pgfplots/xtick = {0,5,...,60},  % make steps of length 5
  xtick={0,2,4,6,8,10,12,14,16,18,20},
  xticklabels={$t_0$,$t_1$,$t_2$,$t_3$,$t_4$,$t_5$,$t_6$,$t_7$,$t_8$,$t_9$,$t_{10}$},
  axis background/.style = {fill=white},
  ylabel = {Transaction Lifecycle Duration (s)},
  xlabel = {Time Steps},
  tick align = outside,
  label style={font=\tiny},
  tick label style={font=\tiny},]
  
\addplot[
    color=blue,
    mark=o,
    mark size=1pt,
    ]
    coordinates {
    (1,4.105)(2,2.719)(3,2.880)(4,2.941)(5,3.304)(6,3.493)(7,4.044)(8,5.518)(9,6.158)(10,6.514)(11,6.367)(12,6.114)(13,6.058)(14,6.553)(15,6.978)(16,5.524)(17,6.169)(18,6.680)(19,7.225)(20,7.632)(21,7.262)
    };
\addplot[
    color=red,
    mark=square,
    mark size=1pt,
    ]
    coordinates {
(1,0.141)(2,1.237)(3,0.673)(4,0.461)(5,0.647)(6,0.699)(7,0.650)(8,0.579)(9,0.737)(10,0.620)(11,0.667)(12,0.711)(13,0.644)(14,0.664)(15,0.674)(16,0.748)(17,0.692)(18,0.677)(19,0.815)(20,0.858)(21,0.741)
    };
\legend{Commit Confirmation,Null Commit Confirmation}
\end{axis} 
\end{tikzpicture}
\caption{Transaction Commit Confirmation Strategy Impact}\label{fig:commit null commit}
  \end{minipage}
\end{figure}

\subsection{CouchDB Document ID \& B-tree Chunk Size}   \label{sec:specialty_couch}
For the following tests we stabilized on a configuration and experimented with document ID generation strategies and CouchDB's B-tree chunk size. 
%and modified the document ID generation strategy
 %(see Section \ref{sec:params} for a description of these parameters). 
%Table \ref{tab:chunk size} reports the results of the chunk size modifications. 
Our configuration included 8 JMeter clients each inserting 54,000 agreement objects, mapped to 4 application servers with 12 workers each. On the Hyperledger Fabric layer, we have a block size of 1200 transactions, 3 second block cut timeout, and 4 endorser peers (1 per application server) out of 6 total peers. 

When an agreement object is generated, the agreement ID is populated with a unique ID (i.e., line 2 in Listing \ref{lst:agr object}). The agreement ID is used to index the agreement JSON object in CouchDB. Table \ref{tab:document id} shows the results of using random and monotonic agreement document IDs.
%Both tests use the same configuration: 8 JMeter clients with 1200 threads and 45 loops, 4 Node.js application servers with 12 workers each, 1200 block size with 3s block cut timeout, 4 endorsing peers (out of 6 total peers), and 16 cpu cores per VM. 
The random ID test uses the \textit{uuid} Node.js library \cite{uuid} to generate the document IDs. Since the IDs are random, inserting and indexing the data object in CouchDB may not leverage the caching behaviour of the underlying B-tree. A random ID data object resulted in 841 TPS, 12.6s latency, and a block fill ratio of 92\%. Alternatively, we leveraged timestamps, which are inherently sequential, as a monotonic document ID. Upon agreement creation, an ID is generated using \textit{process.hrtime} \cite{hrtime}. In order to avoid collisions in document IDs (since multiple agreements could be created nearly simultaneously), the generated timestamp ID is granular (i.e., microseconds). Since monotonic IDs leverage the caching behaviour of the CouchDB B-tree and there are less intermediate tree nodes rewritten, the throughput improved by 5\% to 878 TPS. Average latency is also reduced by 600ms and more blocks were able to be filled (since overall transaction processing improved).

\begin{table}
\captionsetup{font=footnotesize}
\centering
\resizebox{\textwidth}{!}{
\begin{tabular}{|c||c|c|c||c|c||c|c|c|c||c|c|c|}
  \hline
  \multirow{2}{*}{} 
      & \multicolumn{3}{c||}{\textbf{JMeter}}   
          & \multicolumn{2}{|c||}{\textbf{Node.js Server}}            % \cline{2-10}
   		 & \multicolumn{4}{|c||}{\textbf{Hyperledger Fabric}}            % \cline{2-9}
			& \multicolumn{3}{|c|}{\textbf{Results}} \\             \cline{2-13}
  & \textbf{Clts} & \textbf{Thr.} & \textbf{Lps} & \textbf{Srvs} & \textbf{Wrks} & \textbf{BS} & \textbf{TO (s)} & \textbf{End.} & \textbf{ID} & \textbf{Thrpt} & \textbf{Avg. Lat. (ms)} & \textbf{BFR}\\  \hline
  \textbf{TC1} & 8 & 1200 & 45 & 4 & 48 & 1200 & 3 & 4 & Rand & 841 & 12650 & 92\%  \\      \hline
  \textbf{TC2} & 8 & 1200 & 45 & 4 & 48 & 1200 & 3 & 4 & Mono & 878 & 12063 & 94\%  \\      \hline
\end{tabular}}
\caption{CouchDB Document ID Results} 
\label{tab:document id}
\end{table}

\begin{table}
\captionsetup{font=footnotesize}
\centering
\resizebox{\textwidth}{!}{
\begin{tabular}{|c||c|c|c||c|c||c|c|c|c||c|c|c|}
  \hline
  \multirow{7}{*}{} 
      & \multicolumn{3}{c||}{\textbf{JMeter}}   
          & \multicolumn{2}{|c||}{\textbf{Node.js Server}}            % \cline{2-10}
   		 & \multicolumn{4}{|c||}{\textbf{Hyperledger Fabric}}            % \cline{2-9}
			& \multicolumn{3}{|c|}{\textbf{Results}} \\             \cline{2-13}
  & \textbf{Clts} & \textbf{Thr.} & \textbf{Lps} & \textbf{Srvs} & \textbf{Wrks} & \textbf{BS} & \textbf{TO (s)} & \textbf{End.} & \textbf{ChkSz} & \textbf{Thrpt} & \textbf{Avg. Lat. (ms)} & \textbf{BFR}\\  \hline
  \textbf{TC1} & 8 & 1200 & 45 & 4 & 48 & 1200 & 3 & 4 & 256 & 765 & 13468 & 93\%  \\      \hline
  \textbf{TC2} & 8 & 1200 & 45 & 4 & 48 & 1200 & 3 & 4 & 1279 & 769 & 12983 & 81\%  \\      \hline
   \textbf{TC3} & 8 & 1200 & 45 & 4 & 48 & 1200 & 3 & 4 & 2048 & 881 & 11789 & 91\%  \\      \hline
   \textbf{TC4} & 8 & 1200 & 45 & 4 & 48 & 1200 & 3 & 4 & 4096 & 921 & 11525 & 92\%  \\      \hline
 \textbf{TC5} & 8 & 1200 & 45 & 4 & 48 & 1200 & 3 & 4 & 6203 & 891 & 11605 & 94\%  \\      \hline
  \textbf{TC6} & 8 & 1200 & 45 & 4 & 48 & 1200 & 3 & 4 & 8192 & 892 & 11610 & 93\%  \\      \hline
 \textbf{TC7} & 8 & 1200 & 45 & 4 & 48 & 1200 & 3 & 4 & 16384 & 886 & 11882 & 95\%  \\      \hline
\end{tabular}}
\caption{B-tree Chunk Size Results} 
\label{tab:chunk size}
\end{table}

\begin{figure}[t]
  \centering
    \captionsetup{font=footnotesize}
\begin{tikzpicture}
     \begin{axis}[
     width=9cm, height=5cm,  % size of the image
       xmin = 0, xmax = 17,
       ymin = 650, ymax = 950,
       axis y line*=left,
       xlabel={B-tree Chunk Size ($\cdot10^3$)},
       xlabel near ticks,
       ylabel={Throughput (TPS)},
       ylabel near ticks,
      xtick={1,4,8,12,16},
        label style={font=\scriptsize},
  tick label style={font=\scriptsize},
    grid = major,
  grid style = {dashed, gray!30},
legend pos=south east,
legend style={nodes={scale=0.4, transform shape}},
legend columns=2,
     ]
       \addplot[
    color=red,
    mark=square,
    mark size=1pt,
    ]
    coordinates {
(0.256,765)(1.279,769)(2.048,881)(4.096,921)(6.203,891)(8.192,892)(16.384,886)
    }; \label{ppp1}
     \end{axis}
     \begin{axis}[
     width=9cm, height=5cm,  % size of the image
       xmin = 0, xmax = 17,
       ymin = 11, ymax = 14,
       hide x axis,
       axis y line*=right,
       ylabel={Avg Latency (s)},
       ylabel near ticks,
         xtick={0.2,0.6,1,4,8,12,16},
        label style={font=\scriptsize},
  tick label style={font=\scriptsize},
legend pos=south east,
legend style={nodes={scale=0.4, transform shape}},
legend columns=2,
     ]
     \addlegendimage{/pgfplots/refstyle=pp1}\addlegendentry{Throughput}
        \addplot[
    color=blue,
    mark=*,dashed,
    mark size=1pt,
    ]
    coordinates {
(0.256,13.5)(1.279,12.9)(2.048,11.8)(4.096,11.5)(6.203,11.6)(8.192,11.61)(16.384,11.9)
    };  \addlegendentry{Latency}
     \end{axis}
   \end{tikzpicture}
  \caption{Impact of B-tree Chunk Size on Throughput/Latency}\label{fig:throughput_lat_btree}
\end{figure}
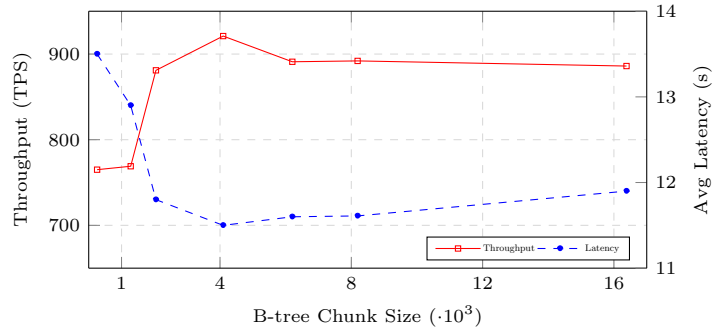

Table \ref{tab:chunk size} reports the results of the chunk size modifications (including monotonic ID). The default B-tree chunk size is 1279, so we experimented with values lower and higher than the default. Test case 1 (TC1) begins with a chunk size of 256, which resulted in 765 TPS. As we increase the chunk size beyond the default value, the throughput improves to 921 TPS with a chunk size of 4096. This is a 20\% throughput increase from the default 1279 value. Latency is also reduced by 1 second with the large chunk size. We observed that chunk size values greater than 4096 start to degrade performance (although the throughput and latency results are still an improvement over the default value results, see Figure \ref{fig:throughput_lat_btree}). For this load and configuration, a chunk size of 4096 optimizes the memory vs. speed trade-off provided by tuning the chunk size. This chunk size is large enough that the number of elements in a leaf node is not too large to hamper insertions, but large enough to avoid unnecessary intermediate node creations (which will slow down insertions). 

Table \ref{tab:couchdb impact} shows the transaction lifecycle breakdown for the key CouchDB configuration tests in Tables \ref{tab:document id} and \ref{tab:chunk size}. Tuning the CouchDB configuration should have a direct impact on processes that interact with the stateDB. Compared to the base case (row 1 of Table \ref{tab:couchdb impact}), the average processing time in CouchDB dropped by 46\% with monotonic IDs (row 2) and a further 11\% with chunk size tuning (row 3). The \textit{CouchDB processing time} metric measures the entire API call (\textit{BatchUpdateDocs}) to CouchDB, which includes the resulting batch update. Therefore, the processing is directly improved by the monotonic ID (i.e., more cache hits) and B-tree chunk size (i.e., less B-tree node rewrites). Since the \textit{average time to commit block changes to the stateDB} includes the BatchUpdateDocs processing time, this metric also improved.
%is also improved, but this metric also includes preparing the batch of transactions and acquiring read/write locks for the database (the time for committing block changes includes the BatchUpdateDocs processing time). 
The \textit{ledger block processing} and \textit{block commit to storage} times increase after monotonic document IDs are introduced because of the increased throughput. With blocks being generated faster, the \textit{average time to validate the Read-Write sets} of transactions in the blocks and the \textit{time to commit the block to the local ledger file} increases. 
%are similar and their differences are likely due to variations in the testing environment.

\begin{table}
\captionsetup{font=footnotesize}
\centering
\resizebox{\textwidth}{!}{
\begin{tabular}{|P{1.5cm}||P{1cm}||P{1.5cm}|P{1.5cm}|P{1.5cm}||P{1cm}|P{1cm}|P{1cm}|P{1.5cm}|P{1.5cm}|P{1cm}||P{1cm}||P{1cm}|}
  \hline
  \multirow{3}{*}{}
      & \multicolumn{1}{c||}{\textbf{Endorse}}   
          & \multicolumn{3}{|c||}{\textbf{Block Creation}}            % \cline{2-10}
   		 & \multicolumn{6}{|c||}{\textbf{Validate \& Commit}}            % \cline{2-9}
			& \multicolumn{1}{|c||}{\textbf{Raft}} 
                      & \multicolumn{1}{|c|}{\textbf{JMeter}} \\             \cline{2-13}
  & \textbf{Avg time to complete proposal (ms)}  & \textbf{Avg block cut time (ms)} & \textbf{Avg time to validate tx (ms)} & \textbf{Avg time to enqueue tx (ms)} & \textbf{Avg time to validate block (ms)} & \textbf{Avg time for ledger block processing (ms)} & \textbf{Avg time to commit block to storage (ms)} & \textbf{Avg time to commit block changes to statedb (ms)} & \textbf{CouchDB processing time - BatchUpdateDocs (ms)} & \textbf{Avg time to commit block (ms)} & \textbf{Raft data persist duration (ms)} & \textbf{Avg latency (ms)}\\  \hline
  \textbf{T~\ref{tab:document id}TC1: base} & 630 & 2097 & 0.87 & 1 & 242 & 149 & 108 & 818 & 220 & 1450  & 4 & 12600  \\      \hline
  \textbf{T~\ref{tab:document id}TC2: monoID} & 561 & 1938 & 0.93 & 2.2 & 275 & 167 & 125 & 813  & 151 & 1462 & 10.9 & 12100  \\      \hline
\textbf{T~\ref{tab:chunk size}TC4: monoID, Chunk Size} & 515 & 1968 & 0.71 & 1.1 & 259 & 161 & 119 & 710  & 136 & 1331 & 9.9 & 11500  \\      \hline
\end{tabular}}
\caption{CouchDB Configuration Impact Results} 
\label{tab:couchdb impact}
\end{table}

This test also illustrates the importance of monitoring the causal relationships between components and protocol steps. Although only the stateDB was tuned, the endorsement phase of the transaction was directly affected by the changes. Compared to the base case, the monotonic IDs and increased chunk size reduced the time spent executing transaction proposals by 22\% (630s to 515s). The reduced CouchDB processing time means the peers will spend less time waiting for CouchDB requests to complete. Since these requests are executing faster, the peers consume less CPU cores interacting with CouchDB. By reducing the CPU consumption in the validate and commit phase, there are more resources for the peers during the endorsement phase. Transaction proposal execution involves CPU intensive operations such as computing signatures (peer endorsement) and chaincode execution, so freeing up resources improves the endorsement phase.

\subsection{State Database}  \label{sec:specialty_statedb} 
As of Hyperledger Fabric v1.4.1, the peer state database options are GolevelDB or CouchDB. GolevelDB is a key-value database embedded in the peers and is enabled by default. CouchDB is an alternative database that runs externally to the peer (e.g., in a separate Docker container). The benefit of using CouchDB is when the chaincode assets (i.e., the agreement data object) are modeled as JSON data \cite{statedb}. Since our application leverages JSON data, we can use rich queries against the chaincode data. However, the benefit of using rich queries comes at a cost of performance. Table \ref{tab:goleveldb} depicts the CouchDB (TC1) and GolevelDB (TC2) state database results.

Both tests use the same configuration with 16 CPU cores per VM. The CouchDB test resulted in 890 TPS and 4.2s latency, whereas GolevelDB provides a throughput boost of 35\% to 1189 TPS and 2.9s latency. A major contributor to the CouchDB performance degradation is the HTTP API (MochiWeb \cite{mochiweb}) through which interactions with the database occur (another contributor is the database locking mechanism, see Section \ref{sec:recommendations}). The reported CouchDB results use batch operations, which write a group of documents to CouchDB through the MochiWeb request handler. However, the use of the MochiWeb API is a performance bottleneck in Hyperledger Fabric, especially with high transactions loads. Since GolevelDB is embedded in the peer, there is no need to interact with the database through an HTTP API, which significantly improves performance. However, there is a functionality and robustness (CouchDB) vs. performance (GolevelDB) trade-off to consider when choosing a state database.

\begin{table}
\captionsetup{font=footnotesize}
\centering
\resizebox{\textwidth}{!}{
\begin{tabular}{|c||c|c|c||c|c||c|c|c|c|c||c|c|c|}
  \hline
  \multirow{2}{*}{} 
      & \multicolumn{3}{c||}{\textbf{JMeter}}   
          & \multicolumn{2}{|c||}{\textbf{Node.js Srv}}            % \cline{2-10}
   		 & \multicolumn{5}{|c||}{\textbf{Hyperledger Fabric}}            % \cline{2-9}
			& \multicolumn{3}{|c|}{\textbf{Results}} \\             \cline{2-14}
  & \textbf{Clts} & \textbf{Thr.} & \textbf{Lps} & \textbf{Srvs} & \textbf{Wrks} & \textbf{BS} & \textbf{TO (s)} & \textbf{End.} & \textbf{ChkSz} & \textbf{DB} & \textbf{Thrpt} & \textbf{Avg. Lat. (ms)} & \textbf{BFR}\\  \hline
  \textbf{TC1} & 6 & 600 & 100 & 2 & 24 & 600 & 2 & 2 & 4096 & CouchDB & 890 & 4221 & 97\%  \\      \hline
  \textbf{TC2} & 6 & 600 & 100 & 2 & 24 & 600 & 2 & 2 & - & GoLevelDB  & 1189 & 2950 & 100\%  \\      \hline
\end{tabular}}
\caption{State Database Choice Results} 
\label{tab:goleveldb}
\end{table}

%\vspace{-1cm}

\subsection{Block Size Impact}  \label{sec:specialty_bs} 

This test reports the impact that the block size (BS) has on each phase of the transaction lifecycle. Tables \ref{tab:block size config} and \ref{tab:block size} show the test configurations and the transaction lifecylce breakdown for 250, 500, and 1000 transaction block sizes, respectively. Figures \ref{fig:peer commit validation} and \ref{fig:orderer cut time} depict the execution times for peer block validation and commit, and orderer block cut time, respectively.  Based on Table \ref{tab:block size config}, these tests were run on VMs with 8 CPU cores, where the 2 application servers with 6 workers each interacted with a Hyperledger Fabric network composed of 4 peers, where 2 of the peers are endorsing transactions. The results show that a block size of 500 provides the optimal throughput of 511 TPS and 6.9s latency, whereas a block size of 1000 begins to degrade performance (decrease in throughput, latency, and block fill ratio).

\begin{table}
\captionsetup{font=footnotesize}
\centering
\resizebox{\textwidth}{!}{
\begin{tabular}{|c||c|c|c||c|c||c|c||c|c||c|c|c|}
  \hline
  \multirow{3}{*}{\textbf{Block Size}} 
      & \multicolumn{3}{c||}{\textbf{JMeter}}   
          & \multicolumn{2}{|c||}{\textbf{Node.js Server}}            % \cline{2-10}
   		 & \multicolumn{2}{|c||}{\textbf{Hyperledger Fabric}}            % \cline{2-9}
	& \multicolumn{2}{|c||}{\textbf{Infra.}}            % \cline{2-9}
			& \multicolumn{3}{|c|}{\textbf{Results}} \\             \cline{2-13}
  & \textbf{Clts} & \textbf{Thr.} & \textbf{Lps} & \textbf{Srvs} & \textbf{Wrks} & \textbf{TO (s)} & \textbf{End.} & \textbf{Cores} & \textbf{Mem (GB)} & \textbf{Thrpt} & \textbf{Avg. Lat. (ms)} & \textbf{BFR} \\  \hline
  \textbf{250} & 6 & 600 & 100 & 2 & 12 & 2 & 2 & 8 & 8 & 478 & 7451 &  99\%  \\      \hline
  \textbf{500} & 6 & 600 & 100 & 2 & 12 & 2 & 2 & 8 & 8 & 511 & 6954 & 97\% \\      \hline
  \textbf{1000} & 6 & 600 & 100 & 2 & 12 & 4 & 2 & 8 & 8 & 506 & 6925 & 96\%  \\      \hline

\end{tabular}}
\caption{Block Size Impact Test Configurations} 
\label{tab:block size config}
\end{table}

Table \ref{tab:block size} helps to determine the source of this degradation. The time a committing peer takes to validate and commit a block, and for the orderer to cut a block all increase by 90-100\% per 250 block size increase. However, the magnitude increases by approximately 100ms for validation, 440ms for block commit, and 500ms for block cut per 250 block size. Therefore, these aspects share the same rate of increase (90-100\%), but the magnitude of increase is different. This shows that the block validation is least affected by the size of the block. The validation process can be run in parallel (\textit{validator pool size} in Section \ref{sec:params}), which means the time to loop through the transactions in a block marginally increases as the number of transactions in a block increases since there are parallel threads validating transactions. Block committing and block cutting do not run in parallel (since only one block can be cut and committed at a time to avoid concurrency issues and chain forking), which results in a much larger time increase (440ms and 500ms, respectively) with bigger block sizes. This also shows that performance optimization should be focused on block commit and block cut since they contribute an order of magnitude more to the overall latency (1000ms combined block commit and cut vs 100ms block validation). For example, selecting the optimal block size for the transaction rate will reduce the average block cut time and tuning the stateDB can reduce the commit processing on the database.

\begin{table}
\captionsetup{font=footnotesize}
\centering
\resizebox{\textwidth}{!}{
\begin{tabular}{|c||P{1cm}||P{1.5cm}|P{1.5cm}|P{1.5cm}||P{1cm}|P{1cm}|P{1cm}|P{1.5cm}|P{1.5cm}|P{1cm}||P{1cm}||P{1cm}|}
  \hline
  \multirow{3}{1cm}{\textbf{Block Size}}
      & \multicolumn{1}{c||}{\textbf{Endorse}}   
          & \multicolumn{3}{|c||}{\textbf{Block Creation}}            % \cline{2-10}
   		 & \multicolumn{6}{|c||}{\textbf{Validate \& Commit}}            % \cline{2-9}
			& \multicolumn{1}{|c||}{\textbf{Raft}} 
                      & \multicolumn{1}{|c|}{\textbf{JMeter}} \\             \cline{2-13}
  & \textbf{Avg time to complete proposal (ms)}  & \textbf{Avg block cut time (ms)} & \textbf{Avg time to validate tx (ms)} & \textbf{Avg time to enqueue tx (ms)} & \textbf{Avg time to validate block (ms)} & \textbf{Avg time for ledger block processing (ms)} & \textbf{Avg time to commit block to storage (ms)} & \textbf{Avg time to commit block changes to statedb (ms)} & \textbf{CouchDB processing time - BatchUpdateDocs (ms)} & \textbf{Avg time to commit block (ms)} & \textbf{Raft data persist duration (ms)} & \textbf{Avg latency (ms)}\\  \hline
  \textbf{250} & 180 & 500 & 1.1 & 1.5 & 107 & 45 & 33	& 285 & 120 	& 497  	& 5.7 & 7451  \\      \hline
  \textbf{500} & 378 & 982 & 1 & 2.1 & 205 & 81 & 62 & 537  & 226 & 936 		& 9.3 & 6954  \\      \hline
\textbf{1000} & 680 & 1986 & 1.1 & 3 & 412 & 152 	& 120 & 1039  & 434 & 1814	& 9.8 & 6925  \\      \hline
\end{tabular}}
\caption{Block Size Impact On Transaction Lifecycle Results.}
\label{tab:block size}
\end{table}

Comparing the 250 block size with the 1000 block size validate and commit results in Figure \ref{fig:peer commit validation} confirms this conclusion. With 250 transactions in a block, the average block commit and validation times (blue line) are relatively close together. However, as the number of transactions in the block grows, the block commit and validations times become farther apart (green line). This shows the importance of tuning the block size since the time to commit the block increases at a higher rate (440ms per 250 block size) than block validation (100ms per 250 block size).

\begin{figure}[!tbp]
  \centering
  %\hspace{-1cm}
  \begin{minipage}[b]{0.45\textwidth}
    \captionsetup{font=footnotesize}
\begin{tikzpicture}
\begin{axis}[
  width=\columnwidth, height=6cm,  % size of the image
  grid = major,
  grid style = {dashed, gray!30},
legend pos=north west,
legend style={nodes={scale=0.3, transform shape}},
legend columns=3,
ylabel near ticks,
  % xmode=log,log basis x=10,
  % ymode=log,log basis y=10,
  xmin = 0,   % start the diagram at this x-coordinate
  xmax = 82,  % end   the diagram at this x-coordinate
  ymin = 0,   % start the diagram at this y-coordinate
  ymax = 2.5, % end   the diagram at this y-coordinate
  %/pgfplots/xtick = {0,5,...,60},  % make steps of length 5
  axis background/.style = {fill=white},
  ylabel = {Execution Time (s)},
  xlabel = {Block Number},
  label style={font=\tiny},
  tick label style={font=\tiny},]
  
  \addplot[color=blue,mark=o,line width=0.5pt,mark size=0.5pt] table[col sep=comma]{Data/Block_Commit_comp_250.csv};
  \addplot[color=red,mark=o,line width=0.5pt,mark size=0.5pt] table[col sep=comma]{Data/Block_Commit_comp_500.csv};
   \addplot[color=green,mark=o,line width=0.5pt,mark size=0.5pt] table[col sep=comma]{Data/Block_Commit_comp_1000.csv};
\addplot[color=blue,mark=square,line width=0.5pt,mark size=0.5pt] table[col sep=comma]{Data/Validate_comp_250.csv};
\addplot[color=red,mark=square,line width=0.5pt,mark size=0.5pt] table[col sep=comma]{Data/Validate_comp_500.csv};
\addplot[color=green,mark=square,line width=0.5pt,mark size=0.5pt] table[col sep=comma]{Data/Validate_comp_1000.csv};
\legend{250 BS (c), 500 BS (c), 1000 BS (c), 250 BS (v), 500 BS (v), 1000 BS (v)}
\end{axis} 
\end{tikzpicture}
\caption{Impact of Block Size on Peer Block Validation (v) and Commit (c)}\label{fig:peer commit validation}
  \end{minipage}
  \hspace{0.5cm}
  %\hfill
  \begin{minipage}[b]{0.45\textwidth}
    \captionsetup{font=footnotesize}
\begin{tikzpicture}
\begin{axis}[
  width=\columnwidth, height=6cm,  % size of the image
  grid = major,
  grid style = {dashed, gray!30},
legend pos=north west,
legend style={nodes={scale=0.6, transform shape}},
ylabel near ticks,
  % xmode=log,log basis x=10,
  % ymode=log,log basis y=10,
  xmin = 0,   % start the diagram at this x-coordinate
  xmax = 82,  % end   the diagram at this x-coordinate
  ymin = 0,   % start the diagram at this y-coordinate
  ymax = 4, % end   the diagram at this y-coordinate
  %/pgfplots/xtick = {0,5,...,60},  % make steps of length 5
  axis background/.style = {fill=white},
  ylabel = {Execution Time (s)},
  xlabel = {Block Number},
  label style={font=\tiny},
  tick label style={font=\tiny},]
  
  \addplot[color=blue,mark=o,line width=0.5pt,mark size=0.5pt] table[col sep=comma]{Data/Block_Cut_comp_250.csv};
  \addplot[color=red,mark=o,line width=0.5pt,mark size=0.5pt] table[col sep=comma]{Data/Block_Cut_comp_500.csv};
  \addplot[color=green,mark=o,line width=0.5pt,mark size=0.5pt] table[col sep=comma]{Data/Block_Cut_comp_1000.csv};
\legend{250 BS, 500 BS, 1000 BS}
\end{axis} 
\end{tikzpicture}
\caption{Impact of Block Size on Orderer Block Cut Time}\label{fig:orderer cut time}
  \end{minipage}
\end{figure}

\subsection{Query Workload Performance Results} \label{sec:query}

The previous test results were based on data insertion workloads. This section provides the results of data query based tests. Query tests fall into two categories based on how many documents are returned in the result set: (i) 1 document is returned; and (ii) 100 documents are returned. The transaction flow for queries differs from insertions as the transactions are not sent to the orderer.

The results of the query tests are shown in Table \ref{tab:query}. The table rows are organized based on the number of documents returned (i.e., 1 or 100). Initially, agreement objects are inserted to the blockchain network and stored in CouchDB. The state database documents are comprised of cached objects for static data (around 500 documents) and agreement objects (1 to 200). For the tests that return 1 result, the database size is just over 500 documents (about 500 cached objects and 1 agreement). Therefore, the test returns the single agreement document from the state database. Tests that return 100 results are run against a database of size 700 (about 500 cached objects and 200 agreements). The test returns 100 out of 200 agreement objects. We also run tests with increasing database sizes of 10,000, 50,000, and 100,000. During all query tests, document types and indexing are enabled.

\begin{table}
\captionsetup{font=footnotesize}
\centering
\resizebox{\textwidth}{!}{
\begin{tabular}{|c||c|c|c||c|c||c|c||c|c||c|c|}
  \hline
  \multirow{14}{*}{} 
      & \multicolumn{3}{c||}{\textbf{JMeter}}   
          & \multicolumn{2}{|c||}{\textbf{Node.js Server}}            % \cline{2-10}
   		 & \multicolumn{2}{|c||}{\textbf{Hyperledger Fabric}}            % \cline{2-9}
	& \multicolumn{2}{|c||}{\textbf{Infra.}}            % \cline{2-9}
			& \multicolumn{2}{|c|}{\textbf{Results}} \\             \cline{2-12}
  & \textbf{Clts} & \textbf{Thr.} & \textbf{Lps} & \textbf{Srvs} & \textbf{Wrks} & \textbf{DB size} & \textbf{ChkSz} & \textbf{Cores} & \textbf{Mem (GB)} & \textbf{Thrpt} & \textbf{Avg. Lat. (ms)} \\  \hline
  \textbf{1} & 1 & 50 & 100 & 1 & 10 & 507 & 4096 & 16 & 16 & 349 & 122   \\      \hline
  \textbf{1} & 1 & 100 & 100 & 1 & 16 & 507 & 4096 & 16 & 16 & 386 & 213   \\      \hline
\textbf{1} & 1 & 150 & 100 & 1 & 16 & 507 & 4096 & 16 & 16 & 410 & 285   \\      \hline
\textbf{1} & 1 & 200 & 100 & 1 & 16 & 507 & 4096 & 16 & 16 & 423 & 411   \\      \hline
\textbf{1} & 1 & 500 & 100 & 1 & 16 & 507 & 4096 & 16 & 16 & 406 & 902   \\      \hline
\textbf{1} & 1 & 1000 & 100 & 1 & 16 & 507 & 4096 & 16 & 16 & 392 & 1865   \\      \hline
\textbf{1} & 1 & 150 & 100 & 1 & 16 & 10507 & 4096 & 16 & 16 & 390 & 303   \\      \hline
\textbf{1} & 1 & 150 & 100 & 1 & 16 & 50507 & 4096 & 16 & 16 & 389 & 335   \\      \hline
\textbf{1} & 1 & 150 & 100 & 1 & 16 & 100507 & 4096 & 16 & 16 & 399 & 325   \\      \hline
  \textbf{100} & 1 & 50 & 100 & 1 & 10 & 707 & 4096 & 16 & 16 & 56 & 870 \\      \hline
\textbf{100} & 1 & 100 & 100 & 1 & 16 & 707 & 4096 & 16 & 16 & 56 & 1782   \\      \hline
\textbf{100} & 1 & 150 & 100 & 1 & 16 & 707 & 4096 & 16 & 16 & 54 & 2736   \\      \hline
\textbf{100} & 1 & 200 & 100 & 1 & 16 & 707 & 4096 & 16 & 16 & 52 & 3787   \\      \hline
\textbf{100} & 1 & 500 & 100 & 1 & 16 & 707 & 4096 & 16 & 16 & 52 & 9353   \\      \hline
\textbf{100} & 1 & 1000 & 100 & 1 & 16 & 707 & 4096 & 16 & 16 & 50 & 20079   \\      \hline
\textbf{100} & 1 & 150 & 100 & 1 & 16 & 10707 & 4096 & 16 & 16 & 43 & 3452   \\      \hline
\textbf{100} & 1 & 150 & 100 & 1 & 16 & 50707 & 4096 & 16 & 16 & 42 & 3547   \\      \hline
\textbf{100} & 1 & 150 & 100 & 1 & 16 & 100707 & 4096 & 16 & 16 & 43 & 3474   \\      \hline
\end{tabular}}
\caption{Query Workload Performance Results} 
\label{tab:query}
\end{table}

Rows 1 to 6 and 10 to 15 of Table \ref{tab:query} show the results of increasing application server workers and client request concurrency (i.e, JMeter load). As expected, returning 1 document is faster than returning 100. Increasing the worker threads on the application server improves the query throughput since more requests can be handled concurrently. As the number of JMeter threads increases (i.e., the amount of query requests), the throughput of the 1 result tests (rows 1-9) increases from 349 TPS to 423 TPS. However, as the number of JMeter threads exceeds 200, the throughput degrades to 392 TPS with 1000 JMeter threads.

Interestingly, the 100 query results have very little throughput degradation with increased request load (starting at 56 TPS and only dropping to 50 TPS). This behaviour can be attributed to the CPU usage of the application server and peer. For the 100 result tests, the peer consumed 60\% of the CPU, whereas the Node.js application consumed 10\%. This means that the application server was underutilized because the peer was busy searching for the 100 documents in CouchDB. Since the application server is slowed down by the peer processing, the throughput is essentially throttled at about 50 TPS. For the 1 result test, the peer cpu usage is at 10\% and the application server is at 50\%. This has the opposite effect from the 100 result tests, where the application server in the 1 result test  is able to process transactions at a much faster rate (since the peer is only querying CouchDB for 1 document) and produce a higher throughput.

The database size can also play a role in performance since the queried document must be searched 
%and indexed 
for in the document space. However, indexing\footnote[3]{The \_\textit{find} API can confirm that indexing is being used \cite{couchFind}} the documents can mitigate the effect that database size has on query performance. Rows 7-9 and 16-18 show the results of large database sizes for 1 and 100 result tests, respectively. Since the documents are indexed, the effect that the database size has on throughput and latency is minimal. The 1 result tests remain around 390 TPS with 300ms latency for database sizes up to 100,000, and the 100 result tests are constant at 42 TPS and 3.5s latency.

%% file: Sections/Cloud_Deployment.tex
\section{Cloud Deployment Performance Results} \label{sec:cloud}

In addition to an on-premise application deployment, we performed initial performance testing with cloud hosted infrastructure. We provide high-level results as the performance optimization methodology applies to a cloud environment and the analysis demonstrated in the previous sections is similar.

\subsection{Infrastructure} The cloud infrastructure for running the performance tests used the IBM Kubernetes Service \cite{kubernetes} that runs on x86 compute. From a cloud perspective, relying on Kubernetes allows for quick provisioning of various cluster topologies, and varying the number of worker nodes, vCPUs, and memory. 
%This infrastructure runs on x86 compute. 
Kubernetes is a container orchestration platform that follows an architectural pattern of master/worker nodes and manages and automates the deployment of containerized workloads. Through deployment strategies, the cluster ensures adequate distribution of resources across all worker nodes. A worker node is similar to a VM and contains services to run pods and is managed by a master node \cite{workernodes}. A pod models a logical host and is a group of tightly-coupled containers with specifications on how to run the containers \cite{pods}. 
The test results below relied on up to 26 worker nodes, each with a single pod (i.e., 10 peers, 3 orderers, 4 Node.js servers, 8 JMeters, TLS certificates), configured with 16 vCPUs, followed by  24 vCPUs.

\subsection{Test Environment} 
\begin{figure}[t]
\captionsetup{font=footnotesize}
  \centering
      \includegraphics[width=0.8\textwidth]{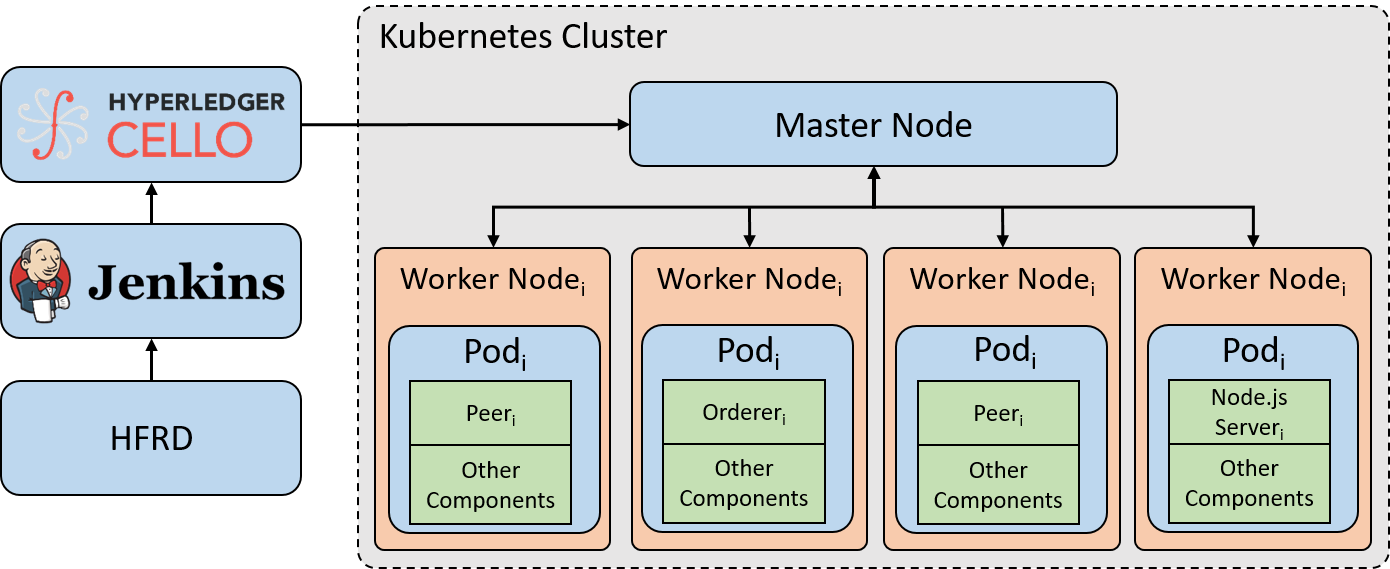}
  \caption{Cloud Deployment Test Environment}
   \label{fig:cloudenv}
\end{figure}

The Hyperledger Fabric Regression Driver (HFRD) performance test tool \cite{HFRD} was responsible for deploying the blockchain components, including the  certificate authorities, peers, and orderers. HFRD is an integration of the Hyperledger Cello Project \cite{cello} with a deployment pipeline controlled by a Jenkins container. The solution provides an administration console to control the environment deployment and the execution of the tests. Although HFRD controls the execution of the overall tests, test scripts (e.g., invocation commands) are executed on the Kubernetes cluster hosting the blockchain network. Due to the limitations of HFRD, the resource allocation of each pod's container is governed by a universal limit. For example, with 16 vCPUs per worker node and the HFRD limit set to 8, the peer and CouchDB containers are deployed in the same pod with 8 vCPUs each (i.e., all containers follow this limit). However, pods running one container (e.g., orderer) are only allocated 8 out of 16 vCPUs. For 24 vCPUs, the limit is set to 12 vCPUs per container. The Node.js application servers and JMeter containers are not deployed with HFRD and are allocated all 16 or 24 vCPUs. The deployed application components were the same as the on-premise deployment. The cloud test environment is shown in Figure \ref{fig:cloudenv}.

\subsection{Results} Table \ref{tab:cloud results} reports the results of 4 performance test cases against an IBM Cloud deployed blockchain application. 
%(same application as the on-premise performance tests). 
Test cases 1 (TC1) and 2 (TC2) show throughput reaching over 1000 TPS with full commit confirmations enabled (Tx Wait). Even with minimal application server scaling (i.e., 2 Node servers) and 16 CPU cores the transaction latency remains reasonable (6.2s) with high throughput (1114 TPS). Scaling the JMeter load, application servers, worker threads, and endorsing peers in TC2 provides an 11\% throughput improvement from TC1, but increases the latency by 3s. The increased latency can be attributed to the processing of over 430,000 transactions in the network. To reduce the latency, we vertically scaled the CPU cores of each worker node to 24 cores in TC3. This improved throughput by 15\% and reduced latency by 1s.

As a final test, TC4 disables the transaction commit confirmation with the \textit{null commit strategy}. This is the same configuration as the on-premise test in Table \ref{tab:null commit} (TC8). In contrast to the on-premise results, the cloud deployment provided a 68\% increase in throughput (1863 vs. 3120 TPS) with a 2s decrease in latency. Comparing the on-premise results to the cloud results yields a subtle difference in testing environments. Based on the dynamic mapping of physical, logical, and virtual CPUs, the on-premise virtualized environment is optimal for average workloads (i.e., not running at full capacity). The cloud infrastructure can be configured in a way that allows the worker nodes to run at full capacity (i.e., high load tests) and optimally utilize the CPU cores with minimal contention between worker nodes. This is especially the case with the null commit strategy test where all Node servers can run at full capacity and provide the large throughput boost.

\begin{table}[t]
\captionsetup{font=footnotesize}
\centering
\resizebox{\textwidth}{!}{
\begin{tabular}{|c||c|c|c||c|c|c||c|c|c||c||c|c|c|}
  \hline
  \multirow{3}{*}{} 
      & \multicolumn{3}{c||}{\textbf{JMeter}}   
          & \multicolumn{3}{|c||}{\textbf{Node.js Server}}            % \cline{2-10}
   		 & \multicolumn{3}{|c||}{\textbf{Hyperledger Fabric}}            % \cline{2-9}
			& \multicolumn{1}{|c||}{\textbf{Infra.}} 
			   & \multicolumn{3}{|c|}{\textbf{Results}} \\             \cline{2-14}
  & \textbf{Clts} & \textbf{Thr.} & \textbf{Lps} & \textbf{Srvs} & \textbf{Wrks} & \textbf{Tx Wait} & \textbf{BS} & \textbf{TO (s)} & \textbf{End.} & \textbf{Cores} & \textbf{Thrpt} & \textbf{Avg. Lat. (ms)} & \textbf{BFR}\\  \hline
  \textbf{TC1} & 6 & 1200 & 25 & 2 & 28 & \cmark & 1200 & 4 & 2 & 16 & 1114 & 6206 & 98\%  \\      \hline
\textbf{TC2} & 8 & 1200 & 45 & 4 & 56 & \cmark & 1200 & 3 & 4 & 16 & 1239 & 9113 & 98\%  \\      \hline
\textbf{TC3} & 8 & 1200 & 45 & 4 & 64 & \cmark & 1400 & 3 & 4 & 24 & 1423 & 8101 & 98\%  \\      \hline
\textbf{TC4} & 8 & 1200 & 70 & 4 & 64 & \xmark & 1200 & 3 & 4 & 16 & 3120 & 2865 & 100\%  \\      \hline
\end{tabular}}
\caption{Cloud Deployment Performance Results} 
\label{tab:cloud results}
\end{table}

%% file: Sections/Improvements.tex
\section{Performance Improvement Recommendations} \label{sec:recommendations}

In this section, we present our recommendations to the application, architecture, and platform (Hyperledger Fabric) layers for performance improvement based on our results and analysis. 

\subsection{State Database Choice} \label{sec:rec_statedb}
The peer state database is currently limited to GolevelDB and CouchDB (v1.4.1). Since many applications rely on JSON data objects, CouchDB is the recommended state database because of rich query support. However, the interface in which the peers interact with CouchDB and the database locking mechanism are bottlenecks in the system. All interactions between the peer and CouchDB are through a REST API, which significantly impacts performance compared to the peer internal GolevelDB. In order to reduce the impact of the REST API, the number of calls to CouchDB should be minimized. Hyperledger Fabric currently uses batch functions to reduce the number of CouchDB calls, however for high transaction rates, the overall number of API calls is still significant. Comparing the use of the REST API for CouchDB with traditional databases is worth investigating.

Database locking mechanisms are necessary for concurrency control, however updating the world state of transactions in CouchDB acquires an exclusive lock on the whole database. Since each transaction execution updates the world state in the stateDB, acquiring a lock on the entire database is a costly operation in terms of lock overhead (the resources used for acquiring and releasing locks) and the lock contention (attempting to acquire a lock held by another process). The granularity of the locking mechanism should be increased in order to reduce the amount of data that the lock is covering. This provides an overhead vs. contention trade-off, where more fine-grained locks consume more resources but reduce lock contention. Integrating PostgreSQL into Hyperledger Fabric has been proposed \cite{postgresql}. PostgreSQL's snapshot isolation levels would allow transaction executions to run in parallel without read and write lock contention \cite{postgresql}. This would significantly improve performance and remove the exclusive lock on the whole database.

As of Hyperledger Fabric v1.4, the query mechanism for CouchDB does not leverage views and map-reduce. Allowing Hyperledger Fabric to utilize CouchDB's views and map-reduce features will improve query performance due to efficient indexing and parallel processing. Future versions may incorporate this feature into Hyperledger Fabric's core APIs.

\subsection{Split Peer Roles} \label{sec:rec_split_peer}
All peers in a Hyperledger Fabric network commit transactions, but there are a subset of peers that also endorse transactions. It is well known that endorsement and commitment are expensive operations \cite{fastfabric,ibmindia}. Endorsing involves 3 steps: (i) checking and validating the proposal; (ii) simulating the proposal; and (iii) endorsing the proposal. First, the transaction proposal headers are validated (i.e, transaction type and signature correctness), the signature is validated (i.e., creator certificate syntax and signature verification), and proposal message verification (i.e., correct chaincode header extensions and payload visibility). The uniqueness of the proposal is confirmed through the transaction ID and the chaincode access control list is checked to see if the proposal complies with the authorized writers. The second step acquires a shared read lock on the stateDB and the proposal is simulated. Proposal simulation executes the chaincode, assembles the Read-Write set, and releases the shared stateDB lock. The last step endorses the proposal (calls the endorsement system chaincode that enforces the endorsement policy and signs the proposed transaction). 

Commitment involves 4 steps: (i) validating the block; (ii) ledger block processing; (iii) committing the block to storage; and (iv) updating the stateDB. First, the block is decoded and all transactions endorsements are validated (signature verification) through the validation system chaincode (VSCC). Then, the state and  read/write sets for all transactions are validated through the multi-version concurrency control (MVCC) checks. Third, metadata is added to the block and the block is committed to the peer's local ledger file. Lastly, the world state based on all transactions is updated in the stateDB.

Endorsing and committing each involve computationally intensive operations, such as signature generations and verifications, marshaling and unmarshaling blocks and transactions, as well as chaincode invocations and stateDB API calls.
Single peers can act as both endorser and committer, which causes increased resource contention on these peers since they must perform transaction endorsement and block commits. However, committing to the stateDB acquires an exclusive lock on the database, which prevents the peers from endorsing transactions since it must read from the stateDB. Splitting the peers  to separate endorser and  committer peer clusters will alleviate the resource contention with the single peer setup. To avoid the database lock contention, endorsement should be fully separated from commitment as suggested in \cite{fastfabric}.

\subsection{Threshold Signatures}  \label{sec:rec_sigs}
Popular consensus protocols such as HotStuff \cite{hotstuff}, Tendermint \cite{tendermint}, and Casper \cite{casper} leverage threshold signatures to improve performance. A $(k, n)$-threshold signature scheme requires $k$ partial signatures from peers out of $n$ total peers to produce a digital signature (the partial signatures are combined). There is a single public key that is held by all peers and each of the $n$ peers has a unique private key. When a digital signature is produced, any peer can verify the signature by using the single public key. Since a subset of peers in Hyperledger Fabric endorse transactions with digital signatures, the resulting endorsement will be composed of a signature per peer. Threshold signatures can be leveraged in peers to reduce storage space (i.e., endorsed transactions will have one combined threshold signature) and improve the performance of block validation (i.e., each transaction will only require validation of the threshold signature rather than a distinct signature from each peer). Threshold signatures have been recommended for peer transaction endorsements \cite{thresholdsigs} and proposed for Hyperledger Fabric \cite{jira}.
%(threshold signatures have not been incorporated as of v1.4.1 of Hyperledger Fabric).

\subsection{Asynchronous Request Design} \label{sec:rec_async}
The \textit{null commit strategy} test in Section \ref{sec:specialty} shows the significant performance improvement when omitting transaction commit confirmations. In order to bridge the gap between the full commit wait strategy and the null commit strategy, we propose that an asynchronous request handling design be integrated in the application tier. This would free up the application server to process more transactions (instead of also consuming resources waiting for commit responses) by having a separate service that handles the commit confirmation for the client. For example, transactions from clients could be stored in a queue, which a ``consumer" service (e.g., Node.js application) connects to. This service would interact with the Hyperledger Fabric network, but hand off the handling of transaction commit confirmation events to a ``listener" service (e.g., Node.js application). 
% that handles the commit confirmation events. 
The listener service notifies the client upon receipt of the transaction confirmation. This design asynchronously handles requests since once the consumer service completes the transaction proposal, it can start processing the next transaction without having transaction and block listeners waiting for commit events. These event listeners are handled in the separate listener service (hosted in a different process).

\subsection{Buffered Channel for Block Validation}  \label{sec:rec_buff}
The block validation routine uses an unbuffered channel for communication between goroutines since it accommodates configuration changes to ACLs and validation rules. However, an unbuffered channel operates synchronously, which means if a goroutine sends its result over the channel, the goroutine will block until a receiving goroutine gets the result from the channel. If many goroutines complete their validation process at the same time then there will be contention between the goroutines to send their results through the channel. Alternatively, a buffered channel allows a capacity to be specified that enables goroutines to ``fire and forget" their validation results. Goroutines will not be blocked when sending and receiving values on the channel as long as the channel is not at full capacity. Leveraging a buffered channel with capacity set to the block size instead of an unbuffered channel may avoid contention between block validation goroutines (Hyperledger Fabric v1.4 does not currently use a buffered results channel), but requires a method to handle configuration changes. We leave experimentally verifying the performance gains of buffered channels as future work.

\subsection{Component Distribution} \label{sec:rec_comp_dist}
Based on our experimental results, the proper distribution of application components is crucial for system performance, especially in a virtualized environment. For non-Kubernetes deployments (e.g., Figure \ref{fig:infra}), an ideal virtualized setup is to have a separate VM for each component to avoid resource and infrastructure limitations. However, for a constrained environment, multiple components can run on a single VM. Assuming all Hyperledger Fabric components run in containers, a peer should run on the same VM as its CouchDB and chaincode instances. The application servers (e.g., Node.js) should run on VMs separate from the peers since the application servers and peers use the most CPU cores. Orderers consume little CPU, so their deployment is not as crucial as other components. For example, they can be placed on VMs that run peers or application servers. Kubernetes-based deployments (e.g., Figure \ref{fig:cloudenv}) should have an anti-affinity rule (i.e., dynamic allocation of worker nodes) that ensures components are properly distributed across resources. Resources can then be constrained to limit CPU and memory allocations to the worker nodes.

%% file: Sections/Conclusion.tex
\section{Conclusion} \label{sec:conlusion}

This paper presented (i) a 
%scientific 
methodology to tune and optimize distributed systems (e.g., Hyperledger Fabric based blockchain applications); (ii) tuning parameters for application and blockchain platform optimizations; (iii) results and analysis from performance testing an enterprise blockchain application; and (iv) recommendations for further performance improvements. The related literature in Section \ref{sec:relatedwork} identified the lack of an in-depth methodical performance tuning strategy for optimizing distributed blockchain-based applications. Furthermore, performance tuning is often reported based on applications that do not satisfy the requirements and features of production systems (e.g., request load, data model and implementation complexity, security). Although these results are beneficial to the community, they are not necessarily transferable to enterprise applications. We aim to bridge the gap between laboratory-based application and production deployed system performance tuning and optimization.

The performance tuning methodology presented in this paper helps analyze distributed systems in the context of enterprise applications. By initially performing a solution assessment and system health check, we can determine the state of the system infrastructure, architecture, and application implementation. With a satisfactory system state, all application components (e.g., application server, blockchain network) are isolated to understand the causal relationships between components and how transactions pass through the system. After knowing where potential bottlenecks are in the system and why these bottlenecks occur, we examine the components for what can be tuned and optimized. Tuning component parameters and application implementation based on request load and infrastructure are paramount for high throughput and low latency. Horizontally and vertically scaling the application architecture and infrastructure has significant performance implications due to increased parallelization and reduced resource contention. We leverage logs, tracing, and metrics to collect application performance data.

We then applied this methodology to optimize a production-grade distributed blockchain application. The performance experiments were run against an application deployed in production that uses production quality code, full logging and transaction tracing support, access control and authentication, data caching, and utilized the Hyperledger Fabric blockchain platform. Our experiments included horizontally scaling application servers and peers, specialty tests (e.g., commit strategies, stateDB, block size),  queries, and on-premise and cloud deployments. Memory and CPU cores were also vertically scaled up to 16 GB and 16 cores (including 24 cores in the cloud deployment), respectively. We analyzed the impact of scaling (e.g., endorsing peers) and configurations (e.g., block size) by capturing data from the three main transaction phases (i.e., endorse, order, commit). We reported test results from an on-premise Z system and IBM Cloud deployment. The application of our performance 
%optimization 
methodology produced on-premise results from 30 to 1900 TPS, and cloud results from 1000 to 3000 TPS. %{\color{red}on prem 30 to 1000 tps, cloud 1000 to 3000.}

Based on our experimental results, we proposed a number of recommendations for further performance improvement. From the transaction phase data collected during the stateDB experiments, the interactions with CouchDB are a bottleneck in the system. Altering the current lock strategy or extending Fabric's stateDB pluggability to allow for databases with a more efficient interface (e.g., PostgreSQL) can alleviate this bottleneck. Splitting peer roles to individual endorsers and committers will reduce the resource contention for dual-role peers. There are many CPU intensive operations, such as digital signature generation and verification, performed by peers during transaction endorsement and validation so leveraging threshold signature schemes can reduce the number of signature verification operations. Experimenting with transaction commit strategies allowed us to determine how an asynchronous request handling design would reduce the idle time waiting for transaction commit confirmations in the current synchronous commit handling model. Analyzing the block verification procedure revealed the use of unbuffered channels for validation goroutines. Buffered channels based on the block size can reduce blocked goroutines from sending their results to the channel. Lastly, the proper distribution of application components across the underlying infrastructure is crucial for system performance.

%We leveraged a Hyperledger Fabric deployment with 2 organizations and 1 channel  
There are a number of directions for extending this work. As more organizations are onboarded to blockchain applications, use cases may require private and confidential transactions between subsets of organizations. To support this, it is important to determine the scalability of the platform on an organization and channel level. Beyond data segregation, channels could also improve scalability and performance since they can be used in the form of data sharding. A possible direction is determining the breakpoint of how many channels can be introduced before performance degrades.  Incorporating and evaluating application design improvements such as asynchronous request handling is an interesting extension for the application tier. Continuing to reduce latency while maintaining high throughput is also necessary as the number of concurrent requests grows. 

In Section \ref{sec:infrastructure}, we mentioned the use of a 6:1 virtual to physical CPU mapping during our experiments. Based on our analysis, a lower virtual to physical CPU mapping is required to fully utilize all CPU cores. For example, assuming there are 12 vCPUs per VM on our Z system deployment, 7 VMs per LPAR results in 84 virtual IFLs per LPAR. If there are 32 IFLs available to the VMs, then there will be an over commitment ratio of approximately 2.6:1. In some cases, a higher over commitment (such as 6:1) can limit a VM from fully utilizing their CPUs, which may affect overall performance. We are currently experimenting with CPU mapping ratios to allow VMs to fully utilize their CPU cores.